\documentclass[journal]{IEEEtran}
\usepackage[colorlinks,bookmarksopen,bookmarksnumbered,citecolor=red,urlcolor=red]{hyperref}
\usepackage{graphicx}
\usepackage{color,mathtools,float,amsmath,amssymb,amsmath,amsthm,anyfontsize,cite}
\usepackage{mathrsfs,epsfig,epstopdf}
\usepackage{caption}
\usepackage{subcaption}
\usepackage{dblfloatfix}
\usepackage{multirow}
\usepackage[normalem]{ulem}
\usepackage{tabularx,booktabs}
\usepackage{textcomp}
\begin{document}

	\title{High-rate Reliable Communication using Multi-hop and Mesh THz/FSO Networks}
\author{
	\IEEEauthorblockN{Praveen Kumar Singya$^{1}$, Behrooz Makki$^{2}$, \IEEEmembership{Senior Member, IEEE}, Antonio D'Errico$^{3}$, \IEEEmembership{Senior Member, OPTICA}, and Mohamed-Slim Alouini$^{4}$, \IEEEmembership{Fellow, IEEE}}
	\thanks{$^{1}$P. K. Singya is with the Electrical and Electronics   Engineering (EEE) Department, Atal Bihari Vajpayee-Indian Institute of Information Technology and Management (ABV-IIITM), Gwalior, 474015, India (e-mail:praveens@iiitm.ac.in)}
	\thanks{$^{2}$B. Makki is with Ericsson Research, Ericsson, 41756 Göteborg, Sweden
		(e-mail:behrooz.makki@ericsson.com)}
	\thanks{$^{3}$D'Errico is with Ericsson Research, Ericsson, 56124 Pisa, Italy (e-mail:antonio.d.errico@ericsson.com)}
	\thanks{$^{4}$M.-S. Alouini are with the Computer, Electrical, and Mathematical Science and Engineering (CEMSE) Division, King Abdullah University	of Science and Technology (KAUST), Thuwal 23955-6900, Saudi Arabia (e-mail:slim.alouini@kaust.edu.sa)}}
\maketitle

%%%=========================================	

	\begin{abstract}
		In this work, we consider multi-hop and mesh hybrid teraHertz/free-space optics (THz/FSO)-based backhaul networks for high data-rate communications. The results are presented for the cases with both out-band integrated access and backhaul (IAB) and non-IAB based communication setups. We consider different deployments of the THz and FSO networks and consider both switching and combining methods	between the hybrid THz/FSO links.
		We study the impact of atmospheric turbulence, atmospheric attenuation, and the pointing error on the FSO communication. The THz communication suffers from small scale fading, path-loss, and the misalignment error. Finally, we evaluate the effects  of atmospheric attenuation/path-loss, pointing/misalignment error, small-scale fading, atmospheric turbulence, number of antennas, number of user equipments, number of hops, and the threshold data-rates on the performance of considered systems. As we show, with different network deployments and switching/combining methods, the hybrid implementation of the THz/FSO links improves the network reliability significantly.
	\end{abstract}


	\maketitle
	%%%=========================================		
	\section{Introduction}
	A tremendous growth in wirelessly connected Internet devices and mobile traffic is expected in beyond fifth generation (5G) communications. For this, along with the large available bandwidth in millimeter wave (mmWave) and beyond communication bands, various solutions are provided to achieve the high data rates with large capacity and coverage. One such solution is network densification \cite{agiwal2016next} by deploying multiple access points in an area. This increases the need for wireless backhauling,  as fiber installation may not be possible or economically viable. Wireless backhaul currently shares a large portion of the backhaul market and is mainly based on well-planned point-to-point communication in the range of up to 80 GHz. This may need improvements from different aspects in 5G and beyond. For instance, alternative technologies such as free space optics (FSO) can be combined with existing networks to boost the reliability. Also, higher, e.g., sub-THz and THz bands provide more bandwidth for backhauling.  { Finally, as defined by 3GPP, integrated access and backhaul (IAB) refers to the networks in which the same spectrum and/or hardware are used for both backhaul communication and access communication to the user equipments (UEs).} 
	IAB provides flexible wireless backhauling as well as the existing cellular services to the UEs in the same node. Therefore, IAB complements the microwave point-to-point backhaul deployments in dense urban and suburban regions. {Typically, wireless backhaul is based on line-of-sight (LoS) connections which makes the deployment challenging specially in dense areas. As a result, multi-hop communication offered by IAB networks can simplify the deployment of the backhaul networks.
	 Further, both in-band and out-band backhaul operations are supported by the IAB. In the in-band IAB (resp. out-band IAB) operation, the backhaul and the access links operate in the same (resp. different) frequency bands \cite{madapatha2020integrated}.} Note that, in 5G NR, IAB has not been standardized for the sub-THz or THz bands. However, while the scopes of 6G are still not clear, due to the demands on high rate backhauling, in 6G the IAB implementation in high bands may be considered.\par
	
	In the literature, various works study the IAB-enabled communication systems \cite{madapatha2020integrated,madapatha2021topology,zhang2021survey}.  Considering multiple hops,  \cite{islam2017integrated,liu2019joint,yin2022routing} study the resource allocation and routing in IAB networks. 
	Also, \cite{teyeb2019integrated} discusses the node placement strategies in multi-hop IAB network. In \cite{lai2020resource}, the downlink sum-rate is maximized by joint resource allocation and node placement strategies. 
	Further, \cite{polese2020integrated} investigates the performance of mmWave IAB networks and validates their feasibility through simulations.
	Then, \cite{gomez2016optimal} investigates the user throughput of a pico-cell based multi-hop mmWave network which is helpful to deploy the IAB networks for network densification. In \cite{kulkarni2017performance,makki2018performance}, the throughput of an IAB network is analyzed by allocating the resources through dynamic time division
	duplexing (TDD). To handle the interference in mobile IAB networks, \cite{monteiro2022tdd} proposes a TDD design.  Then, \cite{monteiro2022paving} proposes various challenges and solutions towards the mobile IAB design. Considering the infinite Poisson point processes, coverage probability of IAB-based  mmWave networks is characterized in \cite{singh2015tractable,saha2019millimeter}. 
	%%================================
	
	%%================================
	For the upcoming 6G communications, (sub-)THz communication is expected to play a role in high data rate wireless backhauling. The THz bands (100 GHz to 10 THz including sub-THz bands) enables highly secured communication with large directional gain, and provides large operational bandwidth \cite{elayan2019terahertz,rajatheva2020scoring}. However, THz communication may be limited to a short operational range (a few hundred meters) due to large path-loss, misalignment error, and blockage in a dense environment. This may limit the (sub-)THz operations in backhaul only, and access communications may be difficult at THz bands.	
	
	In parallel, the demand for high capacity and data rates guide us towards FSO communication which provides a low-cost alternative to optical fiber cables for backhauling. The FSO enables a secured LoS communication in a license-free optical band with colossal bandwidth and is easily deployed \cite{trichili2020roadmap}. {However, the FSO communication is severely affected by the pointing error, atmospheric attenuation, atmospheric turbulence, and weather conditions like fog and smoke, which limits its operations mainly in backhaul applications. 
	On the other hand, rain, path loss, and fading severely affects the performance of the THz link.  Hence, a parallel deployment of the THz and FSO links can provide a robust and highly reliable backhaul solution.} \par
	
{	Considering mmWave bands for the RF link, various works study the hybrid FSO/RF systems, where both the FSO and the RF links are used separately in different hops (sometimes, referred to as mixed FSO/RF) or in parallel in one of the hops.  For instance,
	\cite{singya2021performance,singya2020performance,makki2017performance,makki2017performance1,zedini2020performance,lee2020throughput,xu2020performance,singya2022Haps,hassan2019hybrid} study various dual or multi-hop hybrid FSO/RF systems by considering the FSO and the RF links in separate hops. On the other hand, \cite{douik2016hybrid,sharma2019switching,swaminathan2021haps,gupta2019hard,nath2019impact,althunibat2020secure,rakia2015outage,makki2016performance} investigate the hybrid FSO/RF systems with the parallel deployment of the FSO and the RF links in at-least one hop, and either switching or simultaneous transmission is performed.}
	Note that \cite{singya2021performance,singya2020performance,makki2017performance,makki2017performance1,zedini2020performance,lee2020throughput,xu2020performance,singya2022Haps,hassan2019hybrid,douik2016hybrid,sharma2019switching,swaminathan2021haps,nath2019impact,gupta2019hard,althunibat2020secure,rakia2015outage,makki2016performance} do not study the THz channel characteristics and its performance in the considered communication systems. Further, \cite{boulogeorgos2019error,li2021performance,bhardwaj2021performance} investigate the performance of dual-hop communication systems with at-lease one THz link, however, do not consider the FSO link in any hop.
	{Considering the FSO and the THz links in separate hops, \cite{li2022mixed,alimi2024performance,liang2024performance} analyze the performance of a hybrid FSO/THz system. On the other hand,
\cite{singyaperformance2022,sharma2023performance,bhardwaj2023performance} consider the parallel deployment of THz and FSO links in atleast one hop.
  \cite{singyaperformance2022} considers a dual-hop FSO/THz system, where both the FSO and the THz links are deployed in parallel in backhaul and mmWave in access link, and both the soft and hard switching schemes are studied.
 However, in \cite{sharma2023performance}, a single -hop communication system with  parallel deployment of THz and FSO link is considered and combining scheme is used at the destination.
 Then, in \cite{bhardwaj2023performance}, a dual-hop communication system with parallel deployment of THz/FSO/mmWave links  between base station (BS) to UAV is considered and outage and BER are calculated.}\par
	
{\subsection{Motivation and Contributions:} The 6G communication requires a reliable very high data rate backhaul communication for massive number of internet devices and UEs.
The higher frequency bands, e.g., sub-THz and THz band can support such demands due to more available bandwidth. Further, FSO is also proven to be a promising option. However, such higher frequencies requires network densification. This increases the need for wireless backhauling. Typically, wireless backhaul is based on LoS connections which makes the deployment challenging specially in dense areas. As a result, multi-hop communication offered by IAB networks can simplify the deployment of the backhaul networks. Further, FSO link has its limitations in severe atmospheric turbulence, weather issues, and pointing error conditions. To compensate this, an RF backup link in parallel can be deployed. However, lower RF bands are incapable to support higher data rates, that can be achieved by THz link with multi-antenna configuration.  Therefore, THz and FSO links with parallel deployment guarantee reliable high data rate communication with improved diversity. 
Also,  various imperfections like misalignment/pointing error and fading/atmospheric turbulence, and path loss in both links must be modeled carefully. Accordingly, we consider different multi-hop system setup by considering the LoS/NLoS connectivity of the FSO link for longer distance.  Then, for proper link performance with considerable complexity, proper switching/combining methods must be done carefully. Finally, a mesh setup in multi-hop scenario enables multiple end-to-end (E2E) routes for improved coverage and facilitates load balancing between different routes.
}

\begin{figure*}[t]
		\centering
		\includegraphics[width=5.2in,height=1.2in]{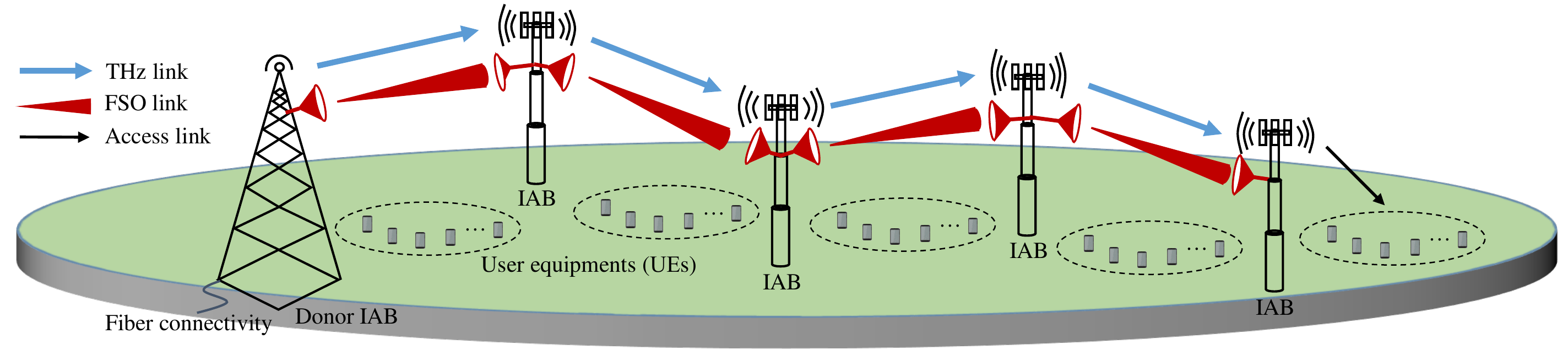}
		\caption{\small{Multi-hop hybrid THz/FSO-based backhaul network with NLoS communication between the macro BS/donor IAB node and the final micro BS/child IAB node.}}
		\label{System1_NLoS}
		\hrulefill
	\end{figure*}

\begin{figure*}[t]
		\centering
		\includegraphics[width=5.2in,height=1.2in]{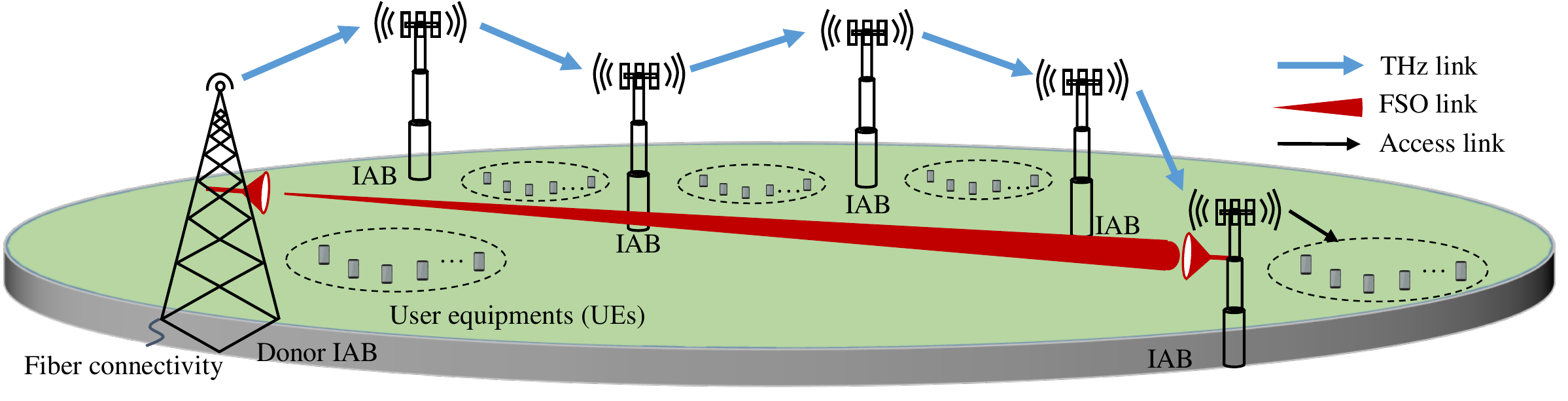}
		\caption{\small{Multi-hop hybrid THz/FSO-based backhaul network with LoS communication between the macro BS/donor IAB node and the final micro BS/child IAB node.}}
		\label{System2_LoS}
		\hrulefill
	\end{figure*}

With above motivation, in this work, we consider mesh and multi-hop hybrid THz/FSO networks for high-rate reliable communications. We present the results for different models of network deployment and for both cases with out-band IAB and non-IAB based communication setups.
	We deploy both the FSO and the THz transceivers at each node according to the network models shown in Figs. \ref{System1_NLoS}-\ref{System2_LoS}. We consider Gamma-Gamma distribution for the FSO link's atmospheric turbulence with Rayleigh distributed pointing error. The $\alpha-\mu$ distribution is considered for the THz link's small-scale fading, along with the Rayleigh distributed pointing error. The THz link is supported with a multi-antenna configuration to reach the same order of rates as in the FSO links. {Further, the successfully decoded information signal at each node is forwarded to the next node through decode-and-forward (DF) relaying. DF is preferred  due to its improved performance over amplified-and-forward (AF) relaying. Moreover, our choice of DF relaying is based on the 3GPP IAB specifications in Releases 15-18 and the fact that, as opposed to AF relaying, DF relaying is less affected by the error propagation effect, which allows for efficient multi-hop or meshed communications. At each node in the multi-hop or the mesh setup, if both the THz and the FSO signals are available, we consider different decoding methods based on either switching between the two signals or their combination. In switching, link with higher SNR is selected. On th other hand, we prefer maximum ratio combining (MRC) to combine the data from both the THz and FSO link at th receiver.} Finally, multiple RF antennas are deployed at each micro BS (MiBS)/child IAB to serve multiple UEs through the mmWave access links, which are Nakagami-m distributed.
	%%===================
	%%====================================
	From this perspective, this study focuses on the following:
	\begin{itemize}
		\item We first obtain closed-form and high signal-to-noise ratio (SNR) approximated probability density function (PDF) and cumulative distribution function (CDF) expressions of the FSO, the THz, and the mmWave access links' SNR.
		
		\item Based on the derived PDF and CDF expressions, we derive the outage probability expressions for both  system models of Figs. \ref{System1_NLoS}-\ref{System2_LoS} as well as the mesh networks with both switching and combining methods between the parallel THz and FSO signals. 
		
		\item Based on the high SNR approximated PDF and CDF expressions, we derive the asymptotic outage probability expressions for the considered system models with switching and combining methods between the parallel THz and FSO signals. Also, we find the diversity order for all considered setups.
		
		\item For both system models, the multi-hop analysis is extended to the mesh network by increasing the number of routes to improve the E2E UE performance. 
		
		\item Finally, we show the impact of atmospheric attenuation/path-loss, pointing/misalignment error, small-scale fading, atmospheric turbulence, number of RF antennas, number of UEs, number of hops, and the threshold data-rates on the performance of considered systems. 
	\end{itemize}  
	From the results, we observe that the parallel deployment of the THz/FSO links in the backhaul with proper system configuration improves the overall performance of the communication system. Also, the diversity order for combining and switching methods remains the same. However, depending on the network deployment, the combining method may provide a considerable performance gain compared to the switching method. Further, higher jitter standard deviation drastically reduces the performance of the THz, the FSO, and consequently the backhaul hybrid THz/FSO link.  Then, the increase in number of hops increases the coverage area, however, reduces the E2E throughput.  Finally, a mesh setup in multi-hop scenario enables multiple E2E routes for improved coverage and facilitates load balancing between different routes. {Finally, to implement a parallel THz/FSO link, different regulatory aspects need to be addressed. Particularly, the co-existence issues of the THz and FSO links with neighbor networks operating at the same or adjacent frequencies should be taken into account. Here, a key point is that the regulatory aspects of THz links are specified by the 3GPP RAN4, while FSO links are not handled by 3GPP. As a result, the regulatory specifications of parallel THz/FSO link require coordination between different organizations.}\par
	
{Note that different from 
	\cite{singya2021performance,singya2020performance,makki2017performance,makki2017performance1,zedini2020performance,lee2020throughput,xu2020performance,singya2022Haps,hassan2019hybrid,boulogeorgos2019error,li2021performance,bhardwaj2021performance,li2022mixed,alimi2024performance,liang2024performance} which have hybrid FSO/RF systems with serial deployment of the FSO and the RF links, we consider parallel deployment of the FSO and THz links. Also, as opposed to 
	 \cite{douik2016hybrid,sharma2019switching,swaminathan2021haps,gupta2019hard,nath2019impact,althunibat2020secure,rakia2015outage,makki2016performance} with parallel deployment of the FSO and RF links, we consider THz links with their channel characteristics and misalignment error. Finally, as compared to \cite{singyaperformance2022} which has a dual-hop FSO/THz system with parallel deployment of the FSO and the THz links with both soft and hard switching, this work focuses on the multi-hop and mesh hybrid THz/FSO networks with parallel deployment as shown in Figs.  \ref{System1_NLoS}-\ref{System2_LoS}. Further, We present the results for both out-band IAB and non-IAB based communication setups  as well as mesh networks which have not
	 been studied before. Conclusively, the considered system models and their analysis makes this work novel from the existing literature.}

	Table \ref{parameter} consists of the notations considered throughout this work. The rest of the paper is organized as follows.
	Section II discusses the considered system and channel models. Then, the performance analysis of System-1 and System-2, shown in Figs. \ref{System1_NLoS}-\ref{System2_LoS}, are discussed in details in Section III and Section IV, respectively. Section V extends the performance of the considered systems to mesh networks. Then, the numerical and simulations results are shown and discussed in Section VI, and finally, the conclusions are drawn in  Section VII.
	{\renewcommand{\arraystretch}{1.3}
		\begin{table*}[]
			%\fontsize{8pt}{8pt}\selectfont
			\caption{{Notations considered throughout the work.}}
			\begin{tabular}{|l|l|l|l|}
				\cline{1-4}
				Notation & Definition & Notation & Definition \\ \cline{1-4}
				$\text{j}\in \{\text{T},\text{F},\text{acc}\}$ & The THz, FSO, or mmWave link's selection parameter &	$n$ & Number of hops \\
				$N_{\text{t}}$ & Transmit antennas at the $n^{\text{th}}$ MiBS/child IAB  &  Q &  Number of routes    \\
				$N_{\text{r}}$ & Antennas  at the $n^{\text{th}}$ MiBS/child IAB THz receiver &  $\kappa$ &  FSO detector type  \\
				$P_{\text{j}_n}$& Transmit power at source j of $n^{\text{th}}$ hop  & 	$L_{\text{j}_n}$ &  The  j link length of $n^{\text{th}}$ hop\\ 
				$G_{\text{t}}^{\text{j}_n}$ & Transmit antenna gain at node j of $n^{\text{th}}$ hop  & $G_{\text{r}}^{\text{j}_n}$& Receive antenna gain at node j of $n^{\text{th}}$ hop\\ 
				$\alpha_n$,$\mu_n$ & $n^{\text{th}}$ $\alpha-\mu$ distributed THz link's fading coefficients &  m & mmWave link's fading parameter\\
				$\alpha_{\text{F}_n}$,$\beta_{\text{F}_n}$&  $n^{\text{th}}$ FSO link's atmospheric turbulence parameters     &  $f_{\text{j}}$   & Operating frequency of the j link \\
				$\xi_\text{j}$,$A_{0,\text{j}}$ & Pointing error parameters for the $\text{j}$ link in $n^{\text{th}}$ hop & $\lambda_\text{j}$   & Wavelength of the j link  \\ 
				$\omega_{\text{j}_n}$,$a_{\text{j}_n}$,$\varepsilon_{\text{j}_n}$ &  Beamwidth, receiver radius, and jitter standard deviation of $\text{j}$ link &  $\varrho_{\text{ox}}$      & Oxygen absorption in mmWave link \\ 
				$h_{{l}_n}$,$I_{{l}_n}$,$p_{{l}_n}$ & Path-loss of the $n^{\text{th}}$  THz, FSO, and mmWave links  & $\varrho_{\text{ra}}$ & Rain attenuation in mmWave link  \\ 
				$\gamma_{\text{j}_n}$ &  Received SNR at node j in $n^{\text{th}}$ hop & c &  Speed of light \\ 
				$\gamma_{\text{th},n}$ & SNR threshold at the $n^{\text{th}}$ FSO and THz receiver   & $\gamma^{\text{UE}}_{\text{th}}$ & SNR threshold at the mmWave UE  \\ 
				\cline{1-4}
			\end{tabular}
			\label{parameter}
	\end{table*}}
	%%%=========================================		
	\section{System and Channel Models}
	We consider multi-hop hybrid THz/FSO networks to provide reliable high data-rates to the mmWave UEs (for the extension of the results to the cases with mesh networks, see Section V). We consider both non-IAB and out-band IAB-based communication setups. With a non-IAB setup, the relay nodes in between do not serve UEs and only forward backhaul data to the end relay node which is responsible for access communication to the UEs. With out-band IAB, on the other hand, while each relay node forwards backhaul data to following IAB nodes, it also provides access links to its surrounding UEs. With an IAB setup, the IAB donor  is the node connecting to the core network
	via, for example, fiber links. Then, the nodes following the IAB donor in a multi-hop fashion are called IAB nodes (see Figs. \ref{System1_NLoS}-\ref{System2_LoS}). With a non-IAB setup, these nodes are referred to as macro BS (MaBS) and micro BS (MiBS), respectively.
	
	For both non-IAB and IAB models, we consider two different network configurations: 
	\begin{itemize}
		\item System 1: The MaBS/donor IAB node and the final MiBS/child IAB node communicate in NLoS manner. In this case, illustrated in Fig. \ref{System1_NLoS}, there are hybrid THz/FSO links in each backhaul link. Such a setup is of interest when there is no strong direct link between the MaBS/donor IAB and the final MiBS/child IAB node.
		
		\item System 2: The MaBS/donor IAB node and the final MiBS/child IAB
		node communicate in LoS manner. In this case, there is a direct FSO link between the MaBS and the final MiBS, as the FSO links can support long ranges. However, because with large hop distances, the THz links may not support the same order of rates as in the FSO links, we do not consider a direct THz link between the MaBS and final MiBS. Instead, in the THz, the data is forwarded to the final MiBS through multiple hops, as illustrated in Fig. \ref{System2_LoS}. 
	\end{itemize}
	
	%	We present the results for the general case with multi-hop and mesh networks. However, it has been previously shown that with multi-hop and IAB networks, the system performance is significantly affected by increasing the number of hops, as the traffic congestion and E2E latency increases \cite{}. For this reason, in practice, the IAB/multi-hop networks are expected to be practically limited to a maximum of two hops.
	Note that System-2 is of interest specially because, although we present the results for the general case with different number of hops, as we explain in the following, in practice multi-hop systems are of interest with a maximum of two hops. In such cases, it is probable to have a strong LoS connection between the first and last nodes, while due to the hop distance, the direct THz connection between these two nodes is not desirable. Moreover, the comparison between System 1 and 2 gives the chance to evaluate the benefits of multi-hop FSO communication.
	
	Both heterodyne detection and intensity modulation/direct detection (IM/DD) methods are considered in the FSO links. Further, we consider DF relaying at each node to forward the successfully decoded signal from the previous node to the next node. For System-1 (resp. System-2) as shown in Fig. \ref{System1_NLoS} (resp. Fig. \ref{System2_LoS}), for each MiBS (resp. for the final MiBS), we consider different techniques for signal reception:
	\begin{itemize}
		\item Switching method: The receiver selects the best signal received through the THz or the FSO links and decodes it.
		\item Combination method: Based on the individual links SNR, the receiver may combine the signals received through the THz and FSO links.
	\end{itemize}
	%%%=============================================================
	\subsection{FSO Link Model}
	Considering the $n^{\text{th}}$ FSO link, the transmitted signal  $s(t)$ received at the MiBS/child IAB\footnote{Note that IAB is an RF technology and the existing IAB systems as defined by 3GPP do not support FSO-based communication.} node is expressed as 	
	\begin{align}
		y_{{\text{F}_n}}\left(t\right)=\left({P_{\text{F}_n}}\eta I_{{l}_n} I_n\right)^{\frac{\kappa}{2}}s(t)+v_{\text{F}_n}\left(t\right),
	\end{align}
	where $P_{\text{F}_n}$ represents the transmit power at the $n^{\text{th}}$ FSO node and $\eta$ is the  optical-to-electrical (O/E) conversion coefficient considered to be the same for all hops. Further, 
	$I_n= I_{\text{p}_n} I_{\text{a}_n}$ represents the fading coefficient of the $n^{\text{th}}$ FSO link, where $I_{\text{p}_n}$ is the pointing error during the transmission and $I_{\text{a}_n}$ represents the fading due to the atmospheric turbulence.  Also, $I_{{l}_n}$ is the atmospheric attenuation of the $n^{\text{th}}$ FSO link. Moreover, $\kappa=1$ and $\kappa=2$ represent the heterodyne detection and IM/DD, respectively, to write the generalized expressions for both the FSO detectors. Furthermore, $v_{\text{F}_n}(t)\sim \mathcal{N}(0, \sigma_{\text{o}}^2)$ represents the additive white Gaussian noise (AWGN) associated with the $n^{\text{th}}$ FSO link with zero mean and $\sigma_{\text{o}}^2$ variance. For simplicity, we consider identical variance for all optical links.  	
	Hence, at the $n^{\text{th}}$ FSO receiver, the instantaneous received SNR  will be $\gamma_{\text{F}_n}={\left(P_{\text{F}_n}\eta I_{{l}_n} I_n\right)^\kappa}/{\sigma^2_{\text{o}}}.$
	
	{\subsubsection{Atmospheric Attenuation} The atmospheric attenuation of the FSO link follows the Beer-Lambert Law \cite{henniger2010introduction} and is characterized as
	\begin{align}
		I_{{l}_n}=\text{exp}\left(-\mathcal{C}_{\text{A}_n}L_{\text{F}_n}\right),
	\end{align}
	where, $L_{\text{F}_n}$ represents the $n^{\text{th}}$ FSO link length. Further, attenuation coefficient $\mathcal{C}_{\text{A}_n}$ is modeled as
	\begin{align}
		\mathcal{C}_{\text{A}_n}=\frac{3.912}{\text{Vi}_n}\left(\frac{\lambda_{\text{F}_n}}{550}\right)^{\text{q}\left(\text{Vi}_n\right)}.
	\end{align}
	Here, for the $n^{\text{th}}$ FSO link, $\text{Vi}_n$ is the visibility with the visibility
	coefficient $\text{q}\left(\text{Vi}_n\right)$, where
	\begin{align}
		\text{q}(\text{Vi}_n)=
		\begin{cases}
			0.585\text{Vi}_n^{1/3},  & \text{for }  \text{Vi}_n < 6 \,\text{km}\\
			1.3,                   & \text{for } 6 < \text{Vi}_n < 50 \,\text{km}\\
			1.6,                   & \text{for } \text{Vi}_n > 50 \,\text{km}.
		\end{cases}
	\end{align}}
	
	\subsubsection{Pointing/Misalignment Error}
\begin{figure}[]
		\centering
		\includegraphics[width=2.7in,height=2in]{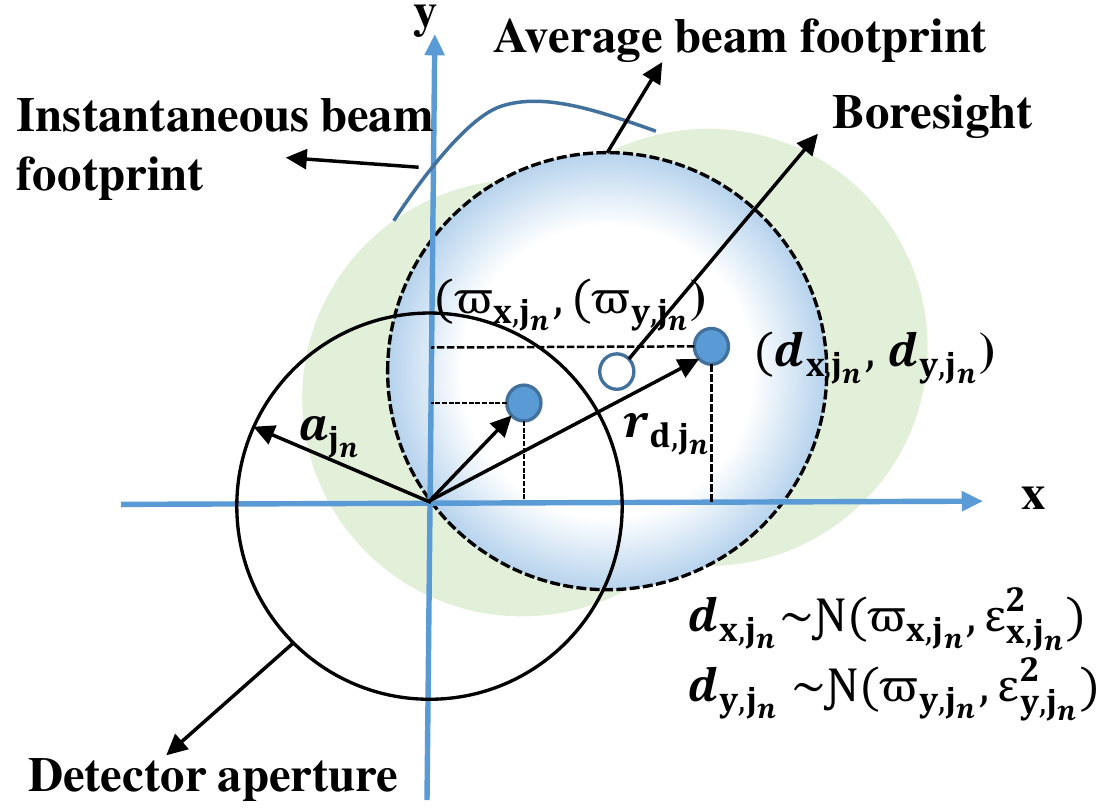}
		\caption{\small{The misalignment/pointing error in the THz/FSO links.}}
		\label{pntfig}
	\end{figure}

	For the $n^{\text{th}}$-hop, we consider an $a_{\text{j}_n}$ aperture radius receiver situated $L_{\text{j}_n}$ distance apart from the transmitter. Hence, at the receiver, the received power from the transmitted $\omega_{\text{j}_n}$ beamwidth Gaussian beam is given as \cite{farid2007outage} 	
	\begin{align}\label{pnt}
		I_{\text{p,j}_n}\left(r_{\text{d},\text{j}_n};L_{\text{j}_n}\right)\approx A_{\text{0,j}_n}~ \text{exp}\left(-{2r_{\text{d,j}_n}^2}/{\omega^2_{\text{eq},\text{j}_n}}\right),
	\end{align}
	where $r_{\text{d,j}_n}$ is the radial displacement between the detector and beam center for the $n^{\text{th}}$-hop. The $A_{\text{0,j}_n}=\left[\text{erf}(v_{0,\text{j}_n})\right]^2$ represents the fraction of the collected power at $r_{\text{d,j}_n}=0$ for the $n^{\text{th}}$-hop, where $v_{\text{0,j}_n}=\sqrt{{a_{\text{j}_n}^2\pi}/{(2\omega_{\text{j}_n}^2)}}$ and $\text{erf}\left(x\right)=\frac{2}{\sqrt{\pi}}\int_{0}^{x}\exp\left(-z^2\right)\text{d}z$ is the error function.
	Further, the equivalent beamwidth is given as $\omega^2_{\text{eq},\text{j}_n}=\frac{\sqrt{A_{\text{0,j}_n}\pi}}{2v_{\text{0,j}_n}~\text{exp}(-v_{\text{0,j}_n}^2)}\omega_{\text{j}_n}^2$.
	Note that for both FSO and THz links, the pointing/misalignment error modeling is considered to be similar and subscript $\text{j} \in [\text{F,T}]$ corresponds to the parameters for the FSO or THz links, respectively. However, the selection parameters may differ for both the THz and FSO links. Figure \ref{pntfig} depicts the pointing/misalignment error for the FSO/THz systems. Here, for the $n^{\text{th}}$-hop,  $d_{\text{x,j}_n}$ is the beam displacement in the horizontal plane and $d_{\text{y,j}_n}$ is the beam displacement in the vertical plane. Consequently, the radial displacement vector is $r_{\text{d,j}_n}=\left[d_{\text{x,j}_n},d_{\text{y,j}_n}\right]^\text{Tr}$.
	%, where $\text{Tr}$ is the transpose of a vector. 
	If $d_{\text{x,j}_n} \approx \mathcal{N}\left(\varpi_{\text{x,j}_n},\varepsilon_{\text{x,j}_n}^2\right)$ and $d_{\text{y,j}_n} \approx \mathcal{N}\left(\varpi_{\text{y,j}_n},\varepsilon_{\text{y,j}_n}^2\right)$ are independently distributed Gaussian random variables (RVs), then $|r_{\text{d,j}_n}|=\sqrt{d_{\text{x,j}_n}^2+d_{\text{y,j}_n}^2}$  will be Beckmann distributed. Hence, different pointing error modeling are possible for different boresight and jitter values \cite{jung2020unified}.
	
	Here, we consider that there is no boresight error, and jitters in horizontal and vertical directions are identical. This corresponds to  $\varpi_{\text{x,j}_n}$ = $\varpi_{\text{y,j}_n}=0$ and $\varepsilon_{\text{x,j}_n}^2=\varepsilon_{\text{y,j}_n}^2=\varepsilon_{\text{j}_n}^2$ for the $n^{\text{th}}$-hop. 
	Thus, the radial displacement is Rayleigh distributed and the PDF of pointing error is 
	\begin{align}\label{Ray_PDF}
		f_{\text{I}_{\text{p,j}_n}}(x)=\frac{\xi_{\text{j}_n}^2}{A_{0,\text{j}_n}^{\xi_{\text{j}_n}^2}}x^{\xi_{\text{j}_n}^2-1},~~~~ \text{for } 0\leq x \leq A_{0,\text{j}_n}
	\end{align}
	where $\xi_{\text{j}_n}={\omega_{\text{eq},\text{j}_n}}/{(2\varepsilon_{\text{j}_n})}$ is a parameter related to the severity of pointing error. 	
	
	%%%%%================================
	\subsubsection{Exact PDF and CDF Expressions of FSO Link} 
	We consider Gamma-Gamma distribution to model the atmospheric turbulence of the $n^{\text{th}}$ FSO link $I_{\text{a}_n}$, which considers the practical weak, moderate, and strong atmospheric turbulence conditions efficiently. The same modeling is also considered in, e.g., \cite{singya2021performance,singyaperformance2022,zedini2016performance,zedini2020performance,swaminathan2021haps,sharma2019switching,makki2016performance,makki2017performance}.  The PDF of the Gamma-Gamma distributed link is defined as \cite{zedini2016performance,zedini2020performance}
	\begin{align}\label{GG}
		f_{I_{\text{a}_n}}(x)&=\frac{2\left(\alpha_{\text{F}_n}\beta_{\text{F}_n}\right)^{\frac{\alpha_{\text{F}_n}+\beta_{\text{F}_n}}{2}}}{\Gamma(\alpha_{\text{F}_n})\Gamma(\beta_{\text{F}_n})}
		x^{\frac{\alpha_{\text{F}_n}+\beta_{\text{F}_n}}{2}-1}\nonumber\\&
		\times\text{K}_{\alpha_{\text{F}_n}-\beta_{\text{F}_n}}\left(2\sqrt{\alpha_{\text{F}_n}\beta_{\text{F}_n} x}\right).
	\end{align}
	 Here, $\Gamma(x)=\int_{0}^{\infty}z^{x-1}\exp\left(-z\right)\text{d}z$ represents complete Gamma function \cite[(6.1.1)]{abramowitz1988handbook} and $\text{K}_{l}(x)=\frac{\Gamma\left(l+\frac{1}{2}\right)}{\sqrt{\pi}}\left(2x\right)^l\int_{0}^{\infty}\frac{\text{cos}(z)}{\left(x^2+z^2\right)^{l+\frac{1}{2}}}\text{d}z$ is the modified Bessel function of second kind with ${l}^{\text{th}}$ order \cite[(9.6.25)]{abramowitz1988handbook}.
	The atmospheric turbulence's fading coefficients for  plane wave propagation are given as \cite{zedini2016performance}
	\begin{align}\label{Atm_tur}
		\alpha_{\text{F}_n}=\left[\text{exp}\left(\frac{0.49\sigma_{\text{RV}_n}^2}{\left(1+1.11\sigma_{\text{RV}_n}^{12/5}\right)^{7/6}}\right)-1\right]^{-1},\nonumber\\
		\beta_{\text{F}_n}=\left[\text{exp}\left(\frac{0.51\sigma_{\text{RV}_n}^2}{\left(1+0.69\sigma_{\text{RV}_n}^{12/5}\right)^{5/6}}\right)-1\right]^{-1},
	\end{align}
	where the Rytov variance $\sigma_{\text{RV}_n}^2=1.23C_{{n}}^2k_{{0}_n}L_{\text{F}_n}^{11/6}$ represents the turbulence strength metric of the $n^{\text{th}}$ FSO link. Further,  $k_{\text{0}_n}=2\pi/\lambda_{\text{F}_n}$ is the wave number and refractive index structure parameter $C_{{n}}^2$ corresponds to the strength of the atmospheric turbulence.

	The fading coefficient $I_n=I_{\text{p}_n}I_{\text{a}_n}$ of the $n^{\text{th}}$ FSO link has the combined  impact of the pointing error and the atmospheric turbulence. Hence, considering zero boresight error with identical jitters for the pointing error and Gamma-Gamma PDF for the atmospheric turbulence, for both detection techniques, the generalized PDF of the $n^{\text{th}}$ FSO link's SNR is given by \cite{zedini2020performance}
	\begin{align}
		f_{\gamma_{\text{F}_n}}(\gamma_{\text{F}_n})&=\frac{\xi^2_{\text{F}_n}}{\kappa\Gamma(\alpha_{\text{F}_n})\Gamma(\beta_{\text{F}_n})\gamma_{\text{F}_n}}\nonumber\\&
		\times\text{G}^{3,0}_{1,3}\left[\frac{\alpha_{\text{F}_n} \beta_{\text{F}_n}}{A_{0{\text{F}_n}}}\left(\frac{\gamma_{\text{F}_n}}{\delta_{{\kappa}_n}}\right)^{\frac{1}{\kappa}}\Big{|}^{\xi^2_{\text{F}_n}+1}_{\xi^2_{\text{F}_n},\alpha_{\text{F}_n},\beta_{\text{F}_n}}\right]
		\label{Gamma_PDF},
	\end{align}
	where 
	$\text{G}^{\hat{m},\hat{n}}_{\hat{p},\hat{q}}\left[z\big{|}^{a_1,...,a_{\hat{n}},a_{\hat{n}+1},...,a_{\hat{p}}}_{b_1,...,b_{\hat{m}},b_{\hat{m}+1},...,b_{\hat{q}}}\right]$ represents the Meijer-G function. Applying $F_{\gamma_{\text{F}_n}}(x)=\int_{0}^{x}f_{\gamma_{\text{F}_n}}(z)\text{d}z$ with \cite[(07.34.21.0084.01)]{wolframe}, the CDF of the  $n^{\text{th}}$ FSO link's SNR  is given by 
	\begin{align}
		F_{\gamma_{\text{F}_n}}(\gamma_{\text{F}_n})&=\mathbb{D}^n_{1}\text{G}^{3\kappa,1}_{\kappa+1,3\kappa+1}\left[\frac{\mathbb{D}^n_{2}\gamma_{\text{F}_n}}{\left(A_{0{\text{F}}_n}\right)^{{\kappa}}\delta_{\kappa_n}}\Big{|}^{1, \Psi^n_{1}}_{\Psi^n_{2}, 0}\right]
		\label{Gamma_CDF},
	\end{align}
	where
	{\small\begin{align}
			\mathbb{D}^n_{1}&=\frac{\kappa^{\left(\alpha_{\text{F}_n}+\beta_{\text{F}_n}-2\right)}\xi_{\text{F}_n}^2}{\left(2\pi\right)^{\kappa-1}\Gamma\left(\alpha_{\text{F}_n}\right)\Gamma\left(\beta_{\text{F}_n}\right)}, 
			\mathbb{D}^n_{2}=\frac{\left(\alpha_{\text{F}_n} \beta_{\text{F}_n}\right)^\kappa}{\kappa^{2\kappa}},\nonumber\\ \Psi^n_{1}&=\left[\frac{\xi_{\text{F}_n}^2+1}{\kappa},...,\frac{\xi_{\text{F}_n}^2+\kappa}{\kappa}\right],	
			\Psi^n_{2}=\Big[\frac{\xi_{\text{F}_n}^2}{\kappa},...,\frac{\xi_{\text{F}_n}^2+\kappa-1}{\kappa},\nonumber\\
			&\frac{\alpha_{\text{F}_n}}{\kappa},...,\frac{\alpha_{\text{F}_n}+\kappa-1}{\kappa},\frac{\beta_{\text{F}_n}}{\kappa},...,\frac{\beta_{\text{F}_n}+\kappa-1}{\kappa}\Big],
	\end{align}}
	Note that the PDF and CDF equations in (\ref{Gamma_PDF}) and (\ref{Gamma_CDF}) are obtained by considering the zero boresight and identical jitter variance while considering the pointing error in FSO link. 
	\subsubsection{High SNR PDF and CDF Expressions  of the FSO Link}
	At high SNRs, the Meijer-G function can be approximated as \cite[ (07.34.06.0001.01)]{wolframe}. Hence,
	the PDF and CDF of the $n^{\text{th}}$ FSO link are approximated as
	\begin{align}
		&f^{A}_{\gamma_{\text{F}_n}}(\gamma_{\text{F}_n})\approx
		\frac{\xi^2_{\text{F}_n}}{\kappa\Gamma(\alpha_{\text{F}_n})\Gamma(\beta_{\text{F}_n})\gamma_{\text{F}_n}}\nonumber\\&
		\times\sum_{p=1}^{3}\left(\frac{\alpha_{\text{F}_n} \beta_{\text{F}_n}}{A_{0{\text{F}_n}}}\left(\frac{\gamma_{\text{F}_n}}{\delta_{{\kappa}_n}}\right)^{\frac{1}{\kappa}}\right)^{\Psi^n_{4,p}}
		\frac{\overset{3}{\underset{\underset{q\neq p}{q=1}}\prod} \Gamma(\Psi^n_{4,q}-\Psi^n_{4,p})}{\overset{1}{\underset{q=1}\prod}\Gamma(\Psi^n_{3,q}-\Psi^n_{4,p})} \label{Asym_FSO_PDF},
	\end{align}
	
	\begin{align}
		&F^\text{A}_{\gamma_{\text{F}_n}}(\gamma_{\text{F}_n})\approx \mathbb{D}^n_{1}\sum_{p=1}^{3\kappa}\left(\frac{\mathbb{D}^n_{2}\gamma_{\text{F}_n}}{(A_{{0\text{F}}_n})^\kappa\delta_{\kappa_n}}\right)^{\Psi^n_{6,p}}\nonumber\\&
		\times	\frac{\overset{3\kappa}{\underset{\underset{q\neq p}{q=1}}\prod} \Gamma(\Psi^n_{6,q}-\Psi^n_{6,p})\overset{1}{\underset{q=1}\prod}\Gamma(1-\Psi^n_{5,q}+\Psi^n_{6,p})}{\overset{\kappa+1}{\underset{q=2}\prod}\Gamma(\Psi^n_{5,q}-\Psi^n_{6,p})\overset{3\kappa+1}{\underset{q=3\kappa+1}\prod}\Gamma(1-\Psi^n_{6,q}+\Psi^n_{6,p})} \label{Asym_FSO_CDF},
	\end{align}
	where, $\Psi^n_{3}=[\xi_{\text{F}_n}^2+1]$, $\Psi^n_{4}=[\xi_{\text{F}_n}^2, \alpha_{\text{F}_n}, \beta_{\text{F}_n}]$, $\Psi^n_{5}=[1,\Psi^n_{1}]$, and $\Psi^n_{6}=[\Psi^n_{2},0]$ for the $n^{\text{th}}$ FSO link. Further,   
	$\Psi^n_{3,p}$/$\Psi^n_{3,q}$, $\Psi^n_{4,p}$/$\Psi^n_{4,q}$, $\Psi^n_{5,p}$/$\Psi^n_{5,q}$, and $\Psi^n_{6,p}$/$\Psi^n_{6,q}$ represent the $p^{\text{th}}$/$q^{\text{th}}$ entry of $\Psi^n_{3}$, $\Psi^n_{4}$, $\Psi^n_{5}$, and $\Psi^n_{6}$, respectively.	
	From (\ref{Asym_FSO_CDF}), the high SNR CDF can be represented as $\approx (G_\text{c}\times \delta_{\kappa_n})^{-G_\text{d}}$ with $G_\text{d}$ and $G_\text{c}$ being the diversity and coding gains, respectively. Here, $G_\text{d}$ can be decided by $\Psi^n_{6}=[\Psi^n_{2},0]=\left[\frac{\xi_{\text{F}_n}^2}{\kappa},...,\frac{\xi_{\text{F}_n}^2+\kappa-1}{\kappa},\frac{\alpha_{\text{F}_n}}{\kappa},...,\frac{\alpha_{\text{F}_n}+\kappa-1}{\kappa},\frac{\beta_{\text{F}_n}}{\kappa},...,\frac{\beta_{\text{F}_n}+\kappa-1}{\kappa},0\right]$. Further, 	$F^\text{A}_{\gamma_{\text{F}_n}}(\gamma_{\text{F}_n})$ has $3\kappa$ summation terms, hence,
	the diversity order of the $n^{\text{th}}$ FSO link is $\min\left(\frac{\xi_{\text{F}_n}^2}{\kappa}, \frac{\alpha_{\text{F}_n}}{\kappa}, \frac{\beta_{\text{F}_n}}{\kappa}\right).$
	
	%%====================================
	\subsection{THz Link}	
	To compensate for the large path-loss at the sub-THz bands, we deploy $N_{\text{r}}$ antennas at the THz receiver with MRC. Considering the $n^{\text{th}}$ THz link, the transmitted signal $s(t)$ received at the MiBS/child IAB node is given by
	\begin{align}		
		y_{\text{T}_n}(t)=\sqrt{P_{\text{T}_n}} h_{{l}_n}\sum_{i=1}^{N_{\text{r}}}h^i_{{\text{T}}_{{n}}}~ s(t)+v_{\text{T}_n}(t),
	\end{align}
	where $P_{\text{T}_n}$ is the transmitted power from the $n^{\text{th}}$ THz source node with $h_{{l}_n}$ being the path-loss.
	Further, for the $n^{\text{th}}$-hop, $h^i_{\text{T}_{n}}=h^i_{\text{f}_n}h^i_{\text{p}_n}$ is the fading coefficient of the THz link to the $i^{\text{th}}$ receiving  antenna that has the combined impact of channel fading $h^i_{\text{f}_n}$ and antenna misalignment $h^i_{\text{p}_n}$. The THz transmitter and receiver are not very far and the receiving antennas are deployed close to each other. Hence, the same pointing error is considered at each receiving antenna. Therefore, the combined channel gain can be written as $||h_{\text{T}_n}||^2=h_{\text{p}_n}^2\sum_{i=1}^{N_\text{r}}|h^i_{\text{f}_n}|^2$. Further, $v_{\text{T}_n}(t)$ is the AWGN associated with the $n^{\text{th}}$ THz link having 0 mean and  $\sigma_{\text{T}}^2$ variance. Note that, for simplicity, the identical AWGN variance $\sigma_{\text{T}}^2$ is considered in each hop.
	Performing MRC at the $n^{\text{th}}$ THz receiver, the instantaneous received SNR  will be
	$\gamma_{\text{T}_n}=\sum_{i=1}^{N_\text{r}}\gamma^i_{\text{T}_n}={P_{\text{T}_n}h_{{l}_n}^2||h_{\text{T}_n}||^2}/{\sigma_{\text{T}}^2}.$
	
	\subsubsection{Path-loss}
	The path-loss $h_{{l}_n}$ of the $n^{\text{th}}$ THz link is expressed as
	\cite{bhardwaj2021performance,kokkoniemi2021line}
	\begin{align}\label{THzPloss}
		&h_{{l}_n}=\frac{c\sqrt{G^{\text{T}_n}_{\text{t}}G^{\text{T}_n}_{\text{r}}}}{4\pi f_{\text{T}_n}d_{\text{T}_n}}
		\nonumber\\&
		\times\exp\left(-\frac{1}{2}d_{\text{T}_n}\left(\sum_{\hat{i}}\mathbb{F}_{\hat{i}}(f_{\text{T}_n},\nu)+\mathbb{G}(f_{\text{T}_n},\nu)\right)\right),
	\end{align}
	where $f_{\text{T}_n}$, $d_{\text{T}_n}$, $G^{\text{T}_n}_{\text{t}}$, and $G^{\text{T}_n}_{\text{r}}$ are the operational frequency, link length, transmit antenna gain, and  receive antenna gain of the $n^{\text{th}}$ THz link, respectively. 
	Here, $\mathbb{G}_{\hat{i}}(f_{\text{T}},\nu)$ is a fitting polynomial to match (\ref{THzPloss}) with the actual response. Further, 
	at 119, 183, 325, 380, 439, and 448 GHz frequencies, the six major absorption lines are shown  by the set of polynomials $\mathbb{F}_{\hat{i}}(f_{\text{T}_n},\nu)$. These parameters are described in \cite[Section II]{singyareport2022}. {Note that, by selecting $\mathbb{G}_{\hat{i}}(f_{\text{T}},\nu)$ for a particular frequency, the THz link can operate till 448 GHz carrier frequency.}
	
	\subsubsection {Exact PDF and CDF expressions  of the THz Link}
	{We consider the experimentally validated $\alpha-\mu$ fading to characterize each of the THz link's gain  \cite{papasotiriou2021experimentally}}. Performing MRC at the THz receiver, the sum of $N_\text{r}$ RVs with $\alpha-\mu$ distributions is also approximated with a single $\alpha-\mu$ RV. Further, the same statistical modeling can be considered for the misalignment error associated with the THz link as shown in Fig. \ref{pntfig}. Hence, we consider the Rayleigh distributed misalignment error for the THz link whose PDF is given in (\ref{Ray_PDF}).

	Considering the combined impact of $\alpha-\mu$ distributed channel fading with Rayleigh distributed misalignment error, 
	%	the PDF of the $n^{\text{th}}$ THz link's channel coefficient is given as 
	%	\begin{align}\label{PDF1}
		%		f_{||h_{\text{T}_n}||}(x)= \mathbb{C}^n_1x^{\xi_{\text{T}_n}^2-1}\Gamma\left(\mathbb{C}^n_2,\mathbb{C}^n_3x^{\alpha_n}\right),
		%	\end{align}
	%	where $\Gamma\left(a,x\right)=\int_{x}^{\infty} z^{a-1}\exp\left(-z\right)\text{d}z$ is the upper incomplete Gamma function. Here, $\mathbb{C}^n_{1}=\frac{\xi^2_{\text{T}_n}}{\left(A_{0,\text{T}_n}\right)^{\xi^2_{\text{T}_n}}}\frac{(N_\text{r}\mu_n)^{\xi^2_{\text{T}_n}/\alpha_n}}{\Omega_n^{\xi^2_{\text{T}_n}}\Gamma(N_\text{r}\mu_n)}$, 
	%	$\mathbb{C}^n_{2}=\frac{\alpha_n N_\text{r}\mu_n-\xi^2_{\text{T}_n}}{\alpha_n}$, $\mathbb{C}^n_{3}=\frac{N_{\text{r}}\mu_n}{\left(A_{{0,\text{T}}_n}\right)^{\alpha_n}\Omega_n^{\alpha_n}}$, and $\Omega_n$ is the $\alpha_n$-root mean value of the fading channel of the $n^{\text{th}}$-hop RVs with $\alpha-\mu$ distributions. Applying $||h_{{\text{T}_n}}||=\sqrt{(\gamma_{{\text{T}_n}}/\bar{\gamma}_{\text{T}_n})}$ in (\ref{PDF1}), 
	the PDF of the SNR of the $n^{\text{th}}$ THz link is obtained as
	{\begin{align}
			f_{\gamma_{\text{T}_n}}(\gamma_{\text{T}_n})&=\frac{\mathbb{C}^n_{1}}{2(\hat{\gamma}_{\text{T}_n})^{\frac{\xi_{\text{T}_n}^2}{2}}}\gamma_{\text{T}_n}^{\frac{\xi_{\text{T}_n}^2}{2}-1}\Gamma\left(\mathbb{C}^n_{2},\mathbb{C}^n_{3}\left(\frac{\gamma_{\text{T}_n}}{\hat{\gamma}_{\text{T}_n}}\right)^{\frac{\alpha_n}{2}}\right)
			\label{PDF_THz},
	\end{align}}
	where $\Gamma\left(a,x\right)=\int_{x}^{\infty} z^{a-1}\exp\left(-z\right)\text{d}z$ is the upper incomplete Gamma function. Here, $\mathbb{C}^n_{1}=\frac{\xi^2_{\text{T}_n}}{\left(A_{0,\text{T}_n}\right)^{\xi^2_{\text{T}_n}}}\frac{(N_\text{r}\mu_n)^{\xi^2_{\text{T}_n}/\alpha_n}}{\Omega_n^{\xi^2_{\text{T}_n}}\Gamma(N_\text{r}\mu_n)}$, 
	$\mathbb{C}^n_{2}=\frac{\alpha_n N_\text{r}\mu_n-\xi^2_{\text{T}_n}}{\alpha_n}$, $\mathbb{C}^n_{3}=\frac{N_{\text{r}}\mu_n}{\left(A_{{0,\text{T}}_n}\right)^{\alpha_n}\Omega_n^{\alpha_n}}$, $\hat{\gamma}_{\text{T}_n}=N_{\text{r}}{\bar{\gamma}}_{\text{T}_n}$, and $\Omega_n$ is the $\alpha_n$-root mean value of the fading channel of the $n^{\text{th}}$-hop RVs with $\alpha-\mu$ distributions. Applying $F_{\gamma_{\text{F}_n}}(x)=\int_{0}^{x}f_{\gamma_{\text{F}_n}}(z)\text{d}z$, CDF of the $n^{\text{th}}$ THz link is derived as 
	
	\begin{align}\label{CDF_THz_N}
		F_{\gamma_{\text{T}_n}}(\gamma_{\text{T}_n})&=\frac{\mathbb{C}^n_{1}}{\xi_{\text{T}_n}^2}\Bigg[\left(\frac{\gamma_{\text{T}_n}}{{\hat{\gamma}}_{\text{T}_n}}\right)^{\frac{\xi_{\text{T}_n}^2}{2}}\Gamma\left(\mathbb{C}^n_{2},\mathbb{C}^n_{3}\left(\frac{\gamma_{\text{T}_n}}{{\hat{\gamma}}_{\text{T}_n}}\right)^{\frac{\alpha_n}{2}}\right)\nonumber\\&
		+{\left(\mathbb{C}^n_{3}\right)}^{-\frac{\xi_{\text{T}_n}^2}{\alpha_n}}\Upsilon\left(N_\text{r}\mu_n,\mathbb{C}^n_{3}\left(\frac{\gamma_{\text{T}_n}}{{\hat{\gamma}}_{\text{T}_n}}\right)^{\frac{\alpha_n}{2}}\right)\Bigg],
	\end{align}	
	where $\Upsilon\left(a,x\right)=\int_{0}^{x} z^{a-1}\exp\left(-z\right)\text{d}z$ is the lower incomplete gamma function.
	
	In Meijer-G form, the CDF of the $n^{\text{th}}$ THz link's SNR  can also be represented as
	\begin{align}\label{CDF_THz}
		F_{\gamma_{\text{T}_n}}(\gamma_{\text{T}_n})&=\frac{\mathbb{C}^n_{1}}{\alpha_n({\hat{\gamma}}_{\text{T}_n})^{\frac{\xi_{\text{T}_n}^2}{2}}}\gamma_{\text{T}_n}^{\frac{\xi_{\text{T}_n}^2}{2}}\nonumber\\&
		\times\text{G}^{2,1}_{2,3}\left[\mathbb{C}^n_{3}\left(\frac{\gamma_{\text{T}_n}}{{\hat{\gamma}}_{\text{T}_n}}\right)^{\frac{\alpha_n}{2}}\Big{|}^{1-\frac{\xi_{\text{T}_n}^2}{\alpha_n},1}_{0,\mathbb{C}^n_{2},-\frac{\xi_{\text{T}_n}^2}{\alpha_n}}\right].
	\end{align}

	\subsubsection{High SNR PDF and CDF expressions of the THz Link}
	Using $\Gamma(m,x)\underset{\overset{x\rightarrow 0}{}}{\approx}\left(\Gamma(m)-\frac{x^m}{m}\right)$ at high SNR, the $n^{\text{th}}$ THz link's PDF (\ref{PDF_THz}) can be approximated as
	{\begin{align}\label{PDF_THz_Asym}
			f^\text{A}_{\gamma_{\text{T}_n}}(\gamma_{\text{T}_n})&\approx\frac{\mathbb{C}^n_{1}}{2({\hat{\gamma}}_{\text{T}_n})^{\frac{\xi_{\text{T}_n}^2}{2}}}\gamma_{\text{T}_n}^{\frac{\xi_{\text{T}_n}^2}{2}-1}\nonumber\\&
			\times\left(\Gamma(\mathbb{C}^n_{2})-\frac{{\left(\mathbb{C}^n_{3}\right)}^{\mathbb{C}^n_{2}}}{\mathbb{C}^n_{2}}\left(\frac{\gamma_{\text{T}}}{{\hat{\gamma}}_{\text{T}_n}}\right)^{\frac{\mathbb{C}^n_{2}\alpha_n}{2}}\right).
	\end{align}	}
	Using  $F^\text{A}_{\gamma_{\text{T}_n}}(\gamma_{\text{T}_n})=\int_{0}^{\gamma_{\text{T}_n}}f^\text{A}_{\gamma_{\text{T}_n}}(z)\text{d}z$, at high SNRs,
	the $n^{\text{th}}$ THz link's CDF is approximated as
	{\begin{align}\label{Asym_THz_CDF}
			F^\text{A}_{\gamma_{\text{T}_n}}(\gamma_{\text{T}_n})&\approx
			\frac{\mathbb{C}^n_{1}\Gamma(\mathbb{C}^n_{2})}{\xi_{\text{T}_n}^2}\left(\frac{\gamma_{\text{T}_n}}{{\hat{\gamma}}_{\text{T}_n}}\right)^{\frac{\xi_{\text{T}_n}^2}{2}}\nonumber\\&
			-
			\frac{\mathbb{C}^n_{1}{\left(\mathbb{C}^n_{3}\right)}^{\mathbb{C}^n_{2}}}{\mathbb{C}^n_{2}\left(\alpha_n N_{\text{r}}\mu_n\right)}\left(\frac{\gamma_{\text{T}_n}}{{\hat{\gamma}}_{\text{T}_n}}\right)^{\frac{\alpha_n N_{\text{r}}\mu_n}{2}}.
	\end{align}	}
	From (\ref{Asym_THz_CDF}), the high SNR CDF can be represented as $\approx(G_\text{c}\times\hat{\gamma}_{\text{T}_n})^{-G_\text{d}}$. Hence, the diversity order of the $n^{\text{th}}$ THz link is  $\min\left(\frac{\xi_{\text{T}_n}^2}{2},\frac{\alpha_n N_{\text{r}}\mu_n}{2}\right)$.
	\subsection{mmWave Access Links}
	We deploy multiple RF antennas ($N_{\text{t}}$) at each MiBS/child IAB node for mmWave transmission to the UEs, which has single antenna due to the small size. Then, with $\hat{s}(t)$ being the successfully decoded signal at the $n^{\text{th}}$ MiBS/child IAB node, the $l^{\text{th}}$ UE receives the signal after maximal ratio transmission (MRT) as
	\begin{align}
		y_{\text{I}_n\text{U}_l}\left(t\right)=\sqrt{P_\text{acc} p_{l}}\sum_{k=1}^{N_\text{t}}h_{{\text{k}_n\text{U}_l}}\hat{s}(t)+v_{\text{U}_l}(t),
	\end{align}
	where $p_{{l}}$ represents the gain of the path-loss in mmWave access link and $P_{\text{acc}}$ is the transmit power at the MiBS/child IAB node. Further, $h_{{\text{k}_n\text{U}_l}}$ is the fading coefficient of the $k^{\text{th}}$ antenna in $n^{\text{th}}$ hop to the $l^{\text{th}}$ UE  and $v_{\text{U}_l}(t)$ is the zero mean and $\sigma_\text{acc}^2$ variance AWGN of the $l^{\text{th}}$ UE's receiver.
	%%	The AWGN noise variance $\sigma_\text{acc}^2$ can be defined as
	%%	$\sigma_\text{acc}^2[\text{dB}]= BW + N_{\text{P}} +N_{\text{f}}$, where $BW$ is the RF receiver bandwidth, $N_{\text{P}}$ is the noise spectral density, and $N_{\text{f}}$ being the noise figure.
	We consider Gamma distributed mmWave access link's gain, i.e.,  $\mathcal{G}(m, \Omega_\text{acc})$, where $m$ is the fading severity and $\Omega_\text{acc}$ is the average power. We consider that all UEs are independent and identically distributed, hence, the same fading severity $m$ and average power is considered for each UE. At the $l^{\text{th}}$ UE, a single Gamma RV  $\approx \mathcal{G}(mN_\text{t}, \Omega_\text{acc})$ can approximate the sum of $N_\text{t}$ Gamma RVs. Hence, the received SNR at $l^{\text{th}}$ UE is  $\gamma_\text{acc}={P_\text{acc}p_{l}||h_\text{acc}||^2}/{\sigma_\text{acc}^2}$ with $||h_\text{acc}||^2=\sum_{k=1}^{N_\text{t}}|h_{{\text{k}_n\text{U}_l}}|^2$ with the following PDF and CDF
	{\begin{align}
			f_{\gamma_{\text{I}_n\text{U}_l}}(\gamma_{\text{acc}})&=\frac{1}{\Gamma\left(mN_\text{t}\right)}\left(\frac{m}{\bar{\gamma}_{\text{acc}}}\right)^{mN_\text{t}}\gamma_{\text{acc}}^{mN_\text{t}-1}\exp{\left(-\frac{m}{\bar{\gamma}_{\text{acc}}}\gamma_{\text{acc}}\right)}\label{PDF_Nak},\\
			~~~F_{\gamma_{\text{I}_n\text{U}_l}}(\gamma_{\text{acc}})&=1-\frac{1}{\Gamma\left(mN_\text{t}\right)}\Gamma\left(mN_\text{t},\frac{m}{\bar{\gamma}_{\text{acc}}}\gamma_{\text{acc}}\right)\label{CDF_Nak},
	\end{align}}
	respectively, where $\bar{\gamma}_{\text{acc}}={\Omega_\text{acc}P_\text{acc}p_{l}}/{\sigma_\text{acc}^2}$ is the average received SNR. 
	
	\subsubsection{Path-loss} The gain of the path-loss $p_{l}$ for the mmWave access link is given as \cite{he2009bit}
	\begin{align}
		p_{{l}}[\text{dB}] = G^\text{acc}_{\text{t}} + G^\text{acc}_{\text{r}}-20\log_{10}\left(\frac{4\pi L_\text{acc}}{\lambda_\text{acc}}\right)-\left(\varrho_{\text{ox}}+ \varrho_{\text{ra}}\right)L_\text{acc},
	\end{align}
	where $G^\text{acc}_\text{t}$, $G^\text{acc}_\text{r}$,  $\lambda_\text{acc}$, and $L_\text{acc}$ are the transmit antenna gain, receiver antenna gain, wavelength, and the link distance of the mmWave access link, respectively. Also, $\varrho_{\text{ra}}$ and $\varrho_{\text{ox}}$  represent respectively the rain attenuation and the oxygen absorption.

	\section{Performance Analysis of System-1}
	For System-1 (as shown in Fig. \ref{System1_NLoS}), we consider the MaBS/donor IAB and the final MiBS/child IAB are in NLoS. Hence, we deploy both the THz and the FSO links in parallel in each hop and the overall backhaul communication is completed through the serial multi-hop hybrid THz/FSO links with DF relaying. Further, we consider both switching and combining methods at the receiver, as explained in the following. We start the analysis for the multi-hop networks. Then, as explained in Section V, the results are extended to the cases with mesh networks.
	%%%================================================
	\subsection {Combining Method}
	\subsubsection{Outage Probability}
	For the considered setup, the THz link is given higher priority and the FSO link is deployed to provide the backup to the THz link, if required. Hence, considering the $n^{\text{th}}$ hybrid THz/FSO backhaul link with combining method, the THz link works alone if its SNR is greater than or equal to an SNR threshold, i.e.,  $\gamma_{\text{T}_n}\geq\gamma_{\text{th},n}$, where $\gamma_{\text{th},n}$ is the threshold. However, with $\gamma_{\text{T}_n}<\gamma_{\text{th},n}$, receiver generates a feedback to activate the FSO link for simultaneous transmission along with the THz link if $\left(\gamma_{\text{T}_n}+\gamma_{\text{F}_n}\right)\geq\gamma_{\text{th},n}$. 
	Therefore, for the $n^{\text{th}}$ backhaul hop, the instantaneous received SNR $\left(\gamma^{\text{CO}}_{\text{I}_n}\right)$ of the hybrid THz/FSO link is given as 
	\begin{align}
		\gamma^{\text{CO}}_{\text{I}_n}=
		\begin{cases}
			\gamma_{\text{T}_n},&\text {for } \gamma_{\text{T}_n}\geq \gamma_{\text{th},n}\\
			\left(\gamma_{\text{T}_n} +\gamma_{\text{F}_n}\right),&\text {for } \gamma_{\text{T}_n} < \gamma_{\text{th},n}~\&~ \left(\gamma_{\text{T}_n}+\gamma_{\text{F}_n}\right)\geq \gamma_{\text{th},n}\\
			0,&\text {for } \left(\gamma_{\text{T}_n}+\gamma_{\text{F}_n}\right)< \gamma_{\text{th},n}.
		\end{cases}
	\end{align}
	
	With combining, outage probability of the $n^{\text{th}}$ hybrid THz/FSO backhaul link is calculated as
	\begin{align}\label{outMRCSC}
		\mathcal{P}^{\text{CO}}_{\gamma_{\text{I}_n}}(\gamma_{\text{th},n})&=F^{\text{CO}}_{\gamma_{\text{I}_n}}(\gamma_{\text{th},n})=\mathcal{P}\left[\left(\gamma_{\text{T}_n}+\gamma_{\text{F}_n}\right)<\gamma_{\text{th},n}\right].
	\end{align}
	Then, (\ref{outMRCSC}) can be derived as
	\begin{align}\label{outMRCmainS1}
		\mathcal{P}^{\text{CO}}_{\gamma_{\text{I}_n}}(\gamma_{\text{th},n})&=
		\mathbb{I}_1(\gamma_{\text{th},n}) - \mathbb{I}_2(\gamma_{\text{th},n}) +\mathbb{I}_3(\gamma_{\text{th},n}).
	\end{align}
	Here,
	\begin{align}\label{outMRC_Closedform}
		\mathbb{I}_1(\gamma_{\text{th},n})&=
		\mathbb{D}^n_{1}\frac{\mathbb{C}^n_{1}\Gamma\left(\mathbb{C}^n_{2}\right)}{\xi_{\text{T}_n}^2{{\bar{\gamma}}_{\text{T}_n}}^{\frac{\xi^2_{\text{T}_n}}{2}}}\Gamma(\mathbb{C}^n_{4})\gamma_{\text{th},n}^{\mathbb{C}^n_{4}-1}\nonumber\\&
		\times\text{G}^{3\kappa,1}_{\kappa+1,3\kappa+1}\left[\frac{\mathbb{D}^n_{2}\gamma_{\text{F}_n}}{\left(A_{0{\text{F}_n}}\right)^{{\kappa}}\delta_{\kappa_n}}\Big{|}^{1, \Psi^n_{1}}_{\Psi^n_{2}, 1-\mathbb{C}^n_{4}}\right],\nonumber\\
		\mathbb{I}_2(\gamma_{\text{th},n})&=\mathbb{D}^n_{1}\frac{\mathbb{C}^n_{1}}{\xi_{\text{T}_n}^2{{\bar{\gamma}}_{\text{T}_n}}^{\frac{\xi^2_{\text{T}_n}}{2}}}\sum_{z_1=0}^{\infty}\frac{(-1)^{z_1}}{z_1!\left(\mathbb{C}^n_{2}+z_1\right)}\left(\frac{\mathbb{C}^n_{3}}{{{\bar{\gamma}}_{\text{T}_n}}^{\frac{\alpha_n}{2}}}\right)^{\mathbb{C}^n_{2}+z_1}\nonumber\\&
		\times\Gamma(\mathbb{C}^n_{5})\gamma_{\text{th},n}^{\mathbb{C}^n_{5}-1}\text{G}^{3\kappa,1}_{\kappa+1,3\kappa+1}\left[\frac{\mathbb{D}^n_{2}\gamma_{\text{F}_n}}{\left(A_{0{\text{F}_n}}\right)^{{\kappa}}\delta_{\kappa_n}}\Big{|}^{1, \Psi^n_{1}}_{\Psi^n_{2}, 1-\mathbb{C}^n_{5}}\right],\nonumber\\
		\mathbb{I}_3(\gamma_{\text{th},n})&=
		\mathbb{D}^n_{1}\frac{\mathbb{C}^n_{1}{\left(\mathbb{C}^n_{3}\right)}^{-\frac{\xi_{\text{T}_n}^2}{\alpha_n}}}{\xi_{\text{T}_n}^2}\sum_{z_2=0}^{\infty}\frac{(-1)^{z_2}}{z_2!\left(N_{\text{r}}\mu_n+z_2\right)}\nonumber\\&
		\times\left(\frac{\mathbb{C}^n_{3}}{{{\bar{\gamma}}_{\text{T}_n}}^{\frac{\alpha_n}{2}}}\right)^{N_{\text{r}}\mu_n+z_2}\Gamma(\mathbb{C}^n_{6})\gamma_{\text{th},n}^{\mathbb{C}^n_{6}-1}\nonumber\\&
		\times
		\text{G}^{3\kappa,1}_{\kappa+1,3\kappa+1}\left[\frac{\mathbb{D}^n_{2}\gamma_{\text{F}_n}}{\left(A_{0{\text{F}_n}}\right)^{{\kappa}}\delta_{\kappa_n}}\Big{|}^{1, \Psi^n_{1}}_{\Psi^n_{2}, 1-\mathbb{C}^n_{6}}\right],
	\end{align} 
	where
	$\mathbb{C}^n_{4}=\frac{\xi^2_{\text{T}_n}}{2}+1$, $\mathbb{C}^n_{5}=\frac{\xi^2_{\text{T}_n}}{2}+\frac{\alpha_n}{2}\left(\mathbb{C}^n_{2}+z_1\right)+1$, and $\mathbb{C}^n_{6}=\frac{\alpha_n}{2}\left(N_{\text{r}}\mu_n+z_1\right)+1$.
	
	Substituting the identities $\mathbb{I}_1(\gamma_{\text{th},n})$, $\mathbb{I}_2(\gamma_{\text{th},n})$, and $\mathbb{I}_3(\gamma_{\text{th},n})$ in (\ref{outMRCmainS1}), outage probability of the $n^{\text{th}}$ hybrid THz/FSO backhaul link for the combining method is obtained.
	
	\textbf{Proof: See Appendix A.}
	
	With DF relaying, outage probability of the $N$-hop hybrid THz/FSO-based backhaul system in a non-IAB setup is obtained as 
	\begin{align}\label{OUT_F_hard}
		\mathcal{P}^{\text{CO}}_{\text{BH,S-1}}&=1-\prod_{\underset{}{n=1}}^{\overset{}{N}}\left(1-F^{\text{CO}}_{\gamma_{\text{I}_n}}\left(\gamma_{\text{th},n}\right)\right).
	\end{align} 
	Also, the E2E outage probability of the considered network at  $l^{\text{th}}$ UE is derived as	
	\begin{align}\label{OUT_F_E2E_MRC}
		\mathcal{P}^{\text{CO}}_{\text{E2E,S-1}}&\left(\gamma^{\text{UE}}_{\text{th}}\right)=1-\left(1-\mathcal{P}^{\text{CO}}_{\text{BH,S-1}}\right)\times\left(1-F_{\gamma_\text{acc}}\left(\gamma^{\text{UE}}_{\text{th}}\right)\right),\nonumber\\&=
		\mathcal{P}^{\text{CO}}_{\text{BH,S-1}}+ F_{\gamma_\text{acc}}\left(\gamma^{\text{UE}}_{\text{th}}\right)-\mathcal{P}^{\text{CO}}_{\text{BH,S-1}}\times F_{\gamma_\text{acc}}\left(\gamma^{\text{UE}}_{\text{th}}\right).
	\end{align}	
	
	\textbf{Note.} The outage probability in (\ref{OUT_F_E2E_MRC}) is for a non-IAB setup. However, with an IAB setup, we need to average the performance over the outage probabilities in each hop to find the overall network outage probability, i.e., 
	with an $N$-hop network, the total outage probability, assuming the same number of UEs per node, is given by 
	\begin{align}\label{IAB_outage}
		\mathcal{P}_{\text{out}}^{\text{total}}=\frac{1}{N}\sum_{n=1}^N \mathcal{P}_{\text{out}}^{n},
	\end{align}
	with $\mathcal{P}_{\text{out}}^{n}$ given by (\ref{OUT_F_E2E_MRC}), except for the case where the donor IAB serves the UEs. In that case, a direct access link exists between the donor and UEs, with outage probability given in (\ref{CDF_Nak}).\par
	
	\textbf{Note.} In an alternative method, one can always use both links simultaneously. In that case, outage probability is given by $\mathcal{P}\left[\left(\gamma_{\text{T}_n}+\gamma_{\text{F}_n}\right)<\gamma_{\text{th},n}\right]$ which ends up the same as in (\ref{outMRCmainS1}).
	
	\subsubsection{Diversity Order for the Combining Method}
	For combining method, the asymptotic outage probability of System-1 is derived as
	\begin{align}\label{Asym_MRC1}
		\mathcal{P}^{\text{CO,A}}_{\gamma_{\text{I}_n}}(\gamma_{\text{th},n})&\approx \int_{0}^{\gamma_{\text{th},n}}f^{\text{A}}_{\gamma_{\text{F}}}(\gamma) \times F^\text{A}_{\gamma_{\text{T}}}(\gamma_{\text{th},n}-\gamma)\text{d}\gamma.
	\end{align}
	This can be simplified as
	\begin{align}\label{Asym_MRC}
		&\mathcal{P}^{\text{CO,A}}_{\gamma_{\text{I}_n}}(\gamma_{\text{th},n})\approx
		\sum_{p=1}^{3} \Theta_1(p)
		\nonumber\\&
		\times\left[\Theta_2\left(\frac{\gamma_{\text{th},n}}{\bar{\gamma}}\right)^{\frac{\xi_{\text{T}_n}^2}{2}+\frac{\Psi_{4,p}}{\kappa}}-
		\Theta_3\left(\frac{\gamma_{\text{th},n}}{\bar{\gamma}}\right)^{\frac{\alpha_n N_{\text{r}}\mu_n}{2}+\frac{\Psi_{4,p}}{\kappa}}\right],
	\end{align}
	where
	$\Theta_1(p)=\frac{\xi^2_{\text{F}_n}}{\kappa\Gamma(\alpha_{\text{F}_n})\Gamma(\beta_{\text{F}_n})}\frac{\overset{3}{\underset{\underset{q\neq p}{q=1}}\prod} \Gamma(\Psi_{4,q}-\Psi_{4,p})}{\overset{1}{\underset{q=1}\prod}\Gamma(\Psi_{3,q}-\Psi_{4,p})}\left(\frac{\alpha_{\text{F}_n} \beta_{\text{F}_n}}{A_{0{\text{F}_n}}\Delta_1^{\frac{1}{\kappa}}}\right)^{\Psi_{4,p}}$, $\Theta_2=\frac{\mathbb{C}^n_{1}\Gamma(\mathbb{C}^n_{2})}{\xi_{\text{T}_n}^2{\left(\Delta_2\right)}^{\frac{\xi_{\text{T}_n}^2}{2}}}\text{B}\left(\frac{\xi_{\text{T}_n}^2}{2}+1,\frac{\Psi_{4,p}}{\kappa}\right)$, and \\
	$\Theta_3=\frac{\mathbb{C}^n_{1}{\left(\mathbb{C}^n_{3}\right)}^{\mathbb{C}^n_{2}}}{\mathbb{C}^n_{2}\left(\alpha_n N_{\text{r}}\mu_n\right){\left(\Delta_2\right)}^{\frac{\alpha_n N_{\text{r}}\mu_n}{2}}}\text{B}\left(\frac{\alpha_n N_{\text{r}}\mu_n}{2}+1,\frac{\Psi_{4,p}}{\kappa}\right).$\par
	\vspace{1em}
	\textbf{Proof: See Appendix B}
	
	From (\ref{Asym_MRC}), the diversity order of the $n^{\text{th}}$ backhaul link with combining is obtained as
	\begin{align}\label{diversity_ST_S1}
		G^{\text{CO},n}_{\text{d}}=\underset{\overset{\text{diversity order of the nth FSO link}}{}}{\underbrace{\min\left(\frac{\xi_{\text{F}_n}^2}{\kappa}, \frac{\alpha_{\text{F}_n}}{\kappa}, \frac{\beta_{\text{F}_n}}{\kappa}\right)}}+\underset{\overset{\text{diversity order of the nth THz link}}{}}{\underbrace{\min\left(\frac{\xi_{\text{T}_n}^2}{2},\frac{\alpha_n N_{\text{r}}\mu_n}{2}\right)}}.
	\end{align}
	Here, the first and the second terms represent the diversity order of the $n^{\text{th}}$ FSO and THz links, respectively. Further, the sum operation represents that both FSO and THz links work in parallel. Also, the diversity order of the $N$-hop hybrid THz/FSO-based backhaul system is given as 
	{\begin{align}\label{diversity_S1_CO}
			&G^{\text{CO}}_{\text{d,S-1}}=\nonumber\\&
			{\underset{\forall_{n=1}^{N}}\min}\left[\underset{\overset{\text{diversity order of the nth hybrid THz/FSO link}}{}}{\underbrace{\min\left(\frac{\xi_{\text{F}_n}^2}{\kappa}, \frac{\alpha_{\text{F}_n}}{\kappa}, \frac{\beta_{\text{F}_n}}{\kappa}\right)+\min\left(\frac{\xi_{\text{T}_n}^2}{2},\frac{\alpha_n N_{\text{r}}\mu_n}{2}\right)}}\right].
	\end{align}}
	Here, the final minimum operation outside the bracket denotes that the diversity order of a multi-hop network is limited by the worst hop.
	Note that the diversity order is the same for the IAB and non-IAB setup because, at high-SNR, the performance is determined by the worst-case scenario which is the outage probability at the end of the multi-hop chain.
	%%==============================================
	\vspace{-1em}
	\subsection{Switching Method}
	\subsubsection{Outage Probability}
	Considering the $n^{\text{th}}$ hybrid THz/FSO backhaul link with switching, the THz link is active, if $\gamma_{\text{T}_n}\geq\gamma_{\text{th},n}$. However, with $\gamma_{\text{T}_n}<\gamma_{\text{th},n}$, receiver generates a feedback  to switch off the THz link and activate the FSO link if $\gamma_{\text{F}_n}\geq\gamma_{\text{th},n}$. 
	Hence, for the $n^{\text{th}}$ hybrid THz/FSO backhaul link, the instantaneous received SNR at the MiBS/IAB $\left(\gamma^{\text{SW}}_{\text{I}_n}\right)$ is 
	\vspace{-0.5em}
	\begin{align}
		\gamma^{\text{SW}}_{\text{I}_n}=
		\begin{cases}
			\gamma_{\text{T}_n}, & \text {for  } \gamma_{\text{T}_n}\geq \gamma_{\text{th},n}\\
			\gamma_{\text{F}_n}, & \text {for  } \gamma_{\text{T}_n} < \gamma_{\text{th},n} ~\&~ \gamma_{\text{F}_n}\geq \gamma_{\text{th},n}\\
			0, & \text {for  } \gamma_{\text{T}_n} < \gamma_{\text{th},n} ~\&~ \gamma_{\text{F}_n}< \gamma_{\text{th},n}.
		\end{cases}
	\end{align}
	
	The hybrid THz/FSO link is in outage when the instantaneous received SNRs of both the THz and the FSO links (i.e., $\gamma_{\text{T}_n}$ and $\gamma_{\text{F}_n}$) are below thresholds.
	Therefore, the outage probability of the $n^{\text{th}}$ hybrid THz/FSO backhaul link  is calculated as
	\begin{align}\label{out11}
		\mathcal{P}^{\text{SW}}_{\gamma_{\text{I}_n}}(\gamma_{\text{th},n})&=F^{\text{SW}}_{\gamma_{\text{I}_n}}(\gamma_{\text{th},n})=\mathcal{P}\left[\gamma_{\text{T}_n}< \gamma_{\text{th},n}~\&~ \gamma_{\text{F}_n}<\gamma_{\text{th},n}\right].
	\end{align}
	The THz and FSO links are statistically independent, hence, outage probability is obtained as
	\begin{align}\label{out12}
		\mathcal{P}^{\text{SW}}_{\gamma_{\text{I}_n}}\left(\gamma_{\text{th},n}\right)&=F_{\gamma_{\text{T}_n}} \left(\gamma_{\text{th},n}\right)\times F_{\gamma_{\text{F}_n}} \left(\gamma_{\text{th},n}\right).
	\end{align}
	
	Substituting  (\ref{Gamma_CDF}) and (\ref{CDF_THz}) in (\ref{out12}), outage probability of the $n^{\text{th}}$ hop is derived as
	{\small\begin{align}\label{outF1}
			\mathcal{P}^{\text{SW}}_{\gamma_{\text{I}_n}}\left(\gamma_{\text{th},n}\right)&=\frac{\mathbb{C}^n_{1}\mathbb{D}^n_{1}}{\alpha_{\text{T}_n}({\bar{\gamma}}_{\text{T}_n})^{\frac{\xi_{\text{T}_n}^2}{2}}}\gamma_{\text{th},n}^{\frac{\xi_{\text{T}_n}^2}{2}}\nonumber\\&
			\times\text{G}^{2,1}_{2,3}\left[\mathbb{C}^n_{3}\left(\frac{\gamma_{\text{th},n}}{{\bar{\gamma}}_{\text{T}_n}}\right)^{\frac{\alpha_{\text{T}_n}}{2}}\Big{|}^{1-\frac{\xi^2_{\text{T}_n}}{\alpha_n},1}_{0,\mathbb{C}^n_{2},-\frac{\xi^2_{\text{T}_n}}{\alpha_n}}\right]\nonumber\\&
			\times\text{G}^{3\kappa,1}_{\kappa+1,3\kappa+1}\left[\frac{\mathbb{D}^n_{2}\gamma_{\text{th},n}}{\left(A_{{0\text{F}_n}}\right)^\kappa\delta_{\kappa_n}}\Big{|}^{1, \Psi^n_{1}}_{\Psi^n_{2}, 0}\right].
	\end{align}}
	
	Also, outage probability of the $N$-hop hybrid THz/FSO-based backhaul system is obtained as 
	\begin{align}\label{OUT_SC_jHop}
		\mathcal{P}^{\text{SW}}_{\text{BH,S-1}}&=1-\overset{N}{\underset{n=1}\prod}\left(1-F^{\text{SW}}_{\gamma_{\text{I}_n}}\left(\gamma_{\text{th},n}\right)\right).
	\end{align}
	
	Finally, the E2E outage probability of System-1 at $l^{\text{th}}$ UE is obtained as	
	\begin{align}\label{OUT_F_E2E_S1}
		\mathcal{P}^{\text{SW}}_{\text{E2E,S-1}}\left(\gamma^{\text{UE}}_{\text{th}}\right)&=1-\left(1-\mathcal{P}^{\text{SW}}_{\text{BH,S-1}}\right)\times\left(1-F_{\gamma_\text{acc}}\left(\gamma^{\text{UE}}_{\text{th}}\right)\right).
	\end{align}	
	\textbf{Note.} Equation (\ref{OUT_F_E2E_S1}) gives the outage probability for the non-IAB setup where the UEs are served only by the end access point. The same as in (\ref{IAB_outage}), the outage probability in the out-band IAB setup is given by averaging the outage probability over all IAB donor and child IAB nodes. 
	%%%================================================
	\subsubsection{Diversity Order for the Switching Method}
	For the $n^{\text{th}}$ hybrid THz/FSO backhaul link, the asymptotic outage probability can be derived as
	\begin{align}\label{Asym_SC1}
		\mathcal{P}^{\text{SW,A}}_{\gamma_{\text{I}_n}}(\gamma_{\text{th},n})&\approx 
		F^\text{A}_{\gamma_{\text{F}}}(\gamma_{\text{th},n}) \times F^\text{A}_{\gamma_{\text{T}}}(\gamma_{\text{th},n}).
	\end{align}
	Substituting $F^\text{A}_{\gamma_{\text{F}}}(\gamma_{\text{th},n})$ and $F^\text{A}_{\gamma_{\text{T}}}(\gamma_{\text{th},n})$ from (\ref{Asym_FSO_CDF}) and (\ref{Asym_THz_CDF}), respectively, in (\ref{Asym_SC1}), we obtain
	{\begin{align}\label{Asym_SC_F1}
			&\mathcal{P}^{\text{SW,A}}_{\gamma_{\text{I}_n}}(\gamma_{\text{th},n})\approx 
			\mathbb{D}^n_1\sum_{p=1}^{3\kappa}
			\left(\frac{\mathbb{D}^n_2\gamma_{\text{th},n}}{(A_{0,\text{F}})^\kappa\delta_{{\kappa}}}\right)^{\Psi^n_{6,p}}
			\nonumber\\&
			\times\frac{\overset{3\kappa}{\underset{\underset{q\neq p}{q=1}}\prod} \Gamma(\Psi^n_{6,q}-\Psi^n_{6,p})\overset{1}{\underset{q=1}\prod}\Gamma(1-\Psi^n_{5,q}+\Psi^n_{6,p})}{\overset{\kappa+1}{\underset{q=2}\prod}\Gamma(\Psi^n_{5,q}-\Psi^n_{6,p})\overset{3\kappa+1}{\underset{q=5\kappa+1}\prod}\Gamma(1-\Psi^n_{6,q}+\Psi^n_{6,p})}\nonumber\\&
			\times\left[\frac{\mathbb{C}^n_{1}\Gamma(\mathbb{C}^n_{2})}{\xi_{\text{T}_n}^2}\left(\frac{\gamma_{\text{th},n}}{{\bar{\gamma}}_{\text{T}_n}}\right)^{\frac{\xi_{\text{T}_n}^2}{2}}-
			\frac{\mathbb{C}^n_{1}{\left(\mathbb{C}^n_{3}\right)}^{\mathbb{C}^n_{2}}}{\mathbb{C}^n_{2}\left(\alpha_n N_{\text{r}}\mu_n\right)}\left(\frac{\gamma_{\text{th},n}}{{\bar{\gamma}}_{\text{T}_n}}\right)^{\frac{\alpha_n N_{\text{r}}\mu_n}{2}}\right].
	\end{align}}
	
	At high SNRs, we have $\bar{\gamma}\rightarrow \infty$. Consequently, the average received SNR of the THz and the FSO links tend to infinity, i.e, $\bar{\gamma}_{\text{T}_n},\delta_{\kappa_n} \rightarrow \infty$. Hence, considering $\bar{\gamma}_{\text{T}_n}=\Delta_1\bar{\gamma}$ and $\delta_{\kappa_n}=\Delta_2\bar{\gamma}$, where $\Delta_1$ and $\Delta_2$ are constants, (\ref{Asym_SC_F1}) can be modified as
	\begin{align}\label{Asym_SC_F}
		\mathcal{P}^{\text{SW,A}}_{\gamma_{\text{I}_n}}(\gamma_{\text{th},n})\approx 
		\sum_{p=1}^{3\kappa}\Theta_4(p)
		\times\bigg[\Theta_5\left(\frac{\gamma_{\text{th},n}}{\bar{\gamma}}\right)^{\Psi^n_{6,p}+\frac{\xi_{\text{T}_n}^2}{2}}\nonumber\\
		-
		\Theta_6\left(\frac{\gamma_{\text{th},n}}{\bar{\gamma}}\right)^{\Psi^n_{6,p}+\frac{\alpha_n N_{\text{r}}\mu_n}{2}}\bigg],
	\end{align}
	where
	$\Theta_4(p)=\mathbb{D}^n_1\frac{\overset{3\kappa}{\underset{\underset{q\neq p}{q=1}}\prod} \Gamma(\Psi^n_{6,q}-\Psi^n_{6,p})\overset{1}{\underset{q=1}\prod}\Gamma(1-\Psi^n_{5,q}+\Psi^n_{6,p})}{\overset{\kappa+1}{\underset{q=2}\prod}\Gamma(\Psi^n_{5,q}-\Psi^n_{6,p})\overset{3\kappa+1}{\underset{q=5\kappa+1}\prod}\Gamma(1-\Psi^n_{6,q}+\Psi^n_{6,p})}	\\
	\left(\frac{\mathbb{D}^n_2}{(A_{0,\text{F}})^\kappa\Delta_1}\right)^{\Psi^n_{6,p}}$, $\Theta_5=\frac{\mathbb{C}^n_{1}\Gamma(\mathbb{C}^n_{2})}{\xi_{\text{T}_n}^2\Delta_2^{\frac{\xi_{\text{T}_n}^2}{2}}}$, and $\Theta_6=\frac{\mathbb{C}^n_{1}{\left(\mathbb{C}^n_{3}\right)}^{\mathbb{C}^n_{2}}}{\mathbb{C}^n_{2}\left(\alpha_n N_{\text{r}}\mu_n\right)\Delta_2^{\frac{\alpha_n N_{\text{r}}\mu_n}{2}}}$.
	
	From (\ref{Asym_SC_F}), the diversity order of the $n^{\text{th}}$ hybrid THz/FSO backhaul link is given as 
	\begin{align}\label{diversity_SW_S1}
		G^{\text{SW},n}_{\text{d}}=\min\left(\frac{\xi_{\text{F}_n}^2}{\kappa}, \frac{\alpha_{\text{F}_n}}{\kappa}, \frac{\beta_{\text{F}_n}}{\kappa}\right)+\min\left(\frac{\xi_{\text{T}_n}^2}{2},\frac{\alpha_n N_{\text{r}}\mu_n}{2}\right).
	\end{align}
	From (\ref{diversity_ST_S1}) and (\ref{diversity_SW_S1}), it is clear that the diversity order of the $n^{\text{th}}$ hybrid THz/FSO backhaul link for both the combining and switching methods is the same. However, combining provides additional SNR gain over switching, which we will see in Section VI. 
	Finally, the diversity order of the $N$-hop hybrid THz/FSO-based backhaul system is given as 
	\begin{align}\label{diversity_S1_SW}
		&G^{\text{SW}}_{\text{d,S-1}}=\nonumber\\&
		\underset{\overset{\forall_{n=1}^{N}}{}}{\min}\left[\underset{\overset{\text{diversity order of the nth hybrid THz/FSO link}}{}}{\underbrace{\min\left(\frac{\xi_{\text{F}_n}^2}{\kappa}, \frac{\alpha_{\text{F}_n}}{\kappa}, \frac{\beta_{\text{F}_n}}{\kappa}\right)+\min\left(\frac{\xi_{\text{T}_n}^2}{2},\frac{\alpha_n N_{\text{r}}\mu_n}{2}\right)}}\right].
	\end{align}
	
	\section{Performance Analysis of System-2}	
	For System-2 (as shown in Fig. \ref{System2_LoS}), we consider the MaBS/donor IAB and the final MiBS/child IAB are in LoS. Hence, the backhaul communication takes place through $N$ cascaded THz links and a single FSO link, with proper combining at the final MiBS/IAB node. We analyze the performance of the proposed System-2 by combining both signals coming from the $N$ cascaded THz links and the single FSO link, or by switching between $N$ cascaded THz links or to the single FSO link for signal reception. Both cases are analyzed in the following.
	
	\subsection{Combining Method}
	\subsubsection{Outage Probability}	Since the FSO link is working as a backup for the cascaded THz links, first we check the availability of THz links in the multi-hop hybrid THz/FSO-based  network. If all THz links experience good channel quality, the equivalent cascaded THz link serves irrespective of the FSO link. If the final THz link is below threshold given that all intermediate THz links experience good channel quality, we consider simultaneous transmission through the cascaded THz links and the FSO link.  Also, if any of the intermediate THz link fails, the FSO link becomes active if $\gamma_{\text{F}} \geq \gamma_{\text{th},N}$.
	Therefore, for the considered System-2 with combining method, the instantaneous received SNR at the final MiBS/child IAB is given as (\ref{S2Com}), at the top of the next page.
	\begin{figure*}
		\begin{align}\label{S2Com}
			\gamma^{\text{CO}}_{\text{CI}}=
			\begin{cases}
				\gamma_{\text{T}_{\text{eq}}}, & \text {for  } \forall_{n=1}^{N}~\gamma_{\text{T}_n} 
				\geq \gamma_{\text{th},n}\\
				\gamma_{\text{T}_n}+\gamma_{\text{F}}, & \text {for  } \forall_{n=1}^{N}\left({\gamma_{\text{T}_n}} < \gamma_{\text{th},n} ~\&~\left(\gamma_{\text{T}_n}+\gamma_{\text{F}}\right) \geq \gamma_{\text{th},N} ~\&~ \gamma_{\text{T}_{n-1}} \geq \gamma_{\text{th},(n-1)}\right) \\
				\gamma_{\text{F}}, & \text {for  } \forall_{n=1}^{N-1}\big(\gamma_{\text{T}_n} < \gamma_{\text{th},n} ~\&~ \gamma_{\text{F}}\geq \gamma_{\text{th},N} ~\&~ \gamma_{\text{T}_{n-1}}\geq \gamma_{\text{th},(n-1)}\big)\\
				0, & \text {for  } \forall_{n=1}^{N-1}\big(\gamma_{\text{T}_n} < \gamma_{\text{th},n} ~\&~ \gamma_{\text{F}}< \gamma_{\text{th},N} ~\&~ \gamma_{\text{T}_{n-1}}\geq \gamma_{\text{th},(n-1)}\big)\\&
				~{\text{or }}\forall_{n=1}^{N}\left(\left(\gamma_{\text{T}_n}+\gamma_{\text{F}}\right) < \gamma_{\text{th},N} ~\&~ \gamma_{\text{T}_{n-1}} \geq \gamma_{\text{th},(n-1)}\right),
			\end{cases}
		\end{align}
	\end{figure*}
	
	Hence, the outage probability of the multi-hop THz/FSO backhaul network is found as
	\begin{align}\label{S2SC1}
		&\mathcal{P}^{\text{CO}}_{\gamma_{\text{CI}}}\nonumber=\\&
		\mathcal{P}\bigg[\forall_{n=1}^{N-1}\big(\gamma_{\text{T}_n} < \gamma_{\text{th},n} ~\&~ \gamma_{\text{F}}< \gamma_{\text{th},N} ~\&~ \gamma_{\text{T}_{n-1}}\geq \gamma_{\text{th},(n-1)}\big)\nonumber\\&
		~{\text{or       }}\forall_{n=1}^{N}\left(\left(\gamma_{\text{T}_n}+\gamma_{\text{F}}\right) < \gamma_{\text{th},N} ~\&~ \gamma_{\text{T}_{n-1}} \geq \gamma_{\text{th},(n-1)}\right)\bigg].	
	\end{align}
	This can be solved as
	\begin{align}\label{S2SC}
		\mathcal{P}^{\text{CO}}_{\gamma_{\text{CI}}}&=
		F_{\gamma_{\text{F}}}\left(\gamma_{\text{th},N}\right)\times\bigg[\sum_{n=1}^{N-1}F_{\gamma_{\text{T}_n}}\left(\gamma_{\text{th},n}\right) \nonumber\\&
		\times \prod_{n_1=1}^{n-1}\left(1-F_{\gamma_{\text{T}_{(n-n_1)}}}\left(\gamma_{\text{th},(n-n_1)}\right)\right)\bigg]\nonumber\\&
		+ \prod_{n=1}^{N-1}\left(1-F_{\gamma_{\text{T}_n}}\left(\gamma_{\text{th},n}\right)\right)\times F^{\text{CO}}_{\gamma_{\text{I}_N}}.
	\end{align}
	Here, $F^{\text{CO}}_{\gamma_{\text{I}_N}}$ is the CDF by performing the combining  between the $N^{\text{th}}$ THz link and the single FSO link at the final node which is calculated the same as in (\ref{outMRCSC}). Further, $F_{\gamma_{\text{T}_n}}\left(\gamma_{\text{th},n}\right)$ is the CDF of the $n^{\text{th}}$ THz link (\ref{CDF_THz}) and $F_{\gamma_{\text{F}}}\left(\gamma_{\text{th},N}\right)$ is the CDF of the FSO link (\ref{Gamma_CDF}).
	Substituting (\ref{Gamma_CDF}), (\ref{CDF_THz}), and (\ref{outMRCSC})  in (\ref{S2SC}),  outage probability of backhaul links in System-2 is obtained.
	
	Finally, the E2E outage probability of the considered network at $l^{\text{th}}$ UE is derived as
	\begin{align}\label{OUT_F_E2E_MRC_S2}
		\mathcal{P}^{\text{CO}}_{\text{E2E,S-2}}\left(\gamma^{\text{UE}}_{\text{th}}\right)&=1-\left(1-\mathcal{P}^{\text{CO}}_{\gamma_{\text{CI}}}\right)\times\left(1-F_{\gamma_\text{acc}}\left(\gamma^{\text{UE}}_{\text{th}}\right)\right).
	\end{align}	
	Note that (\ref{OUT_F_E2E_MRC_S2}) gives the E2E outage probability for the cases with non-IAB based communication setup. Following (\ref{IAB_outage}), one can extend the results to the cases with IAB setup.
	
	Finally, in an alternative method, one can always use both links simultaneously. In that case, the outage probability is given by $\mathcal{P}\Big[\forall_{n=1}^{N-1}\big(\gamma_{\text{T}_n} < \gamma_{\text{th},n} ~\&~ \gamma_{\text{F}}< \gamma_{\text{th},N} ~\&~ \gamma_{\text{T}_{n-1}}\geq \gamma_{\text{th},(n-1)}\big)
	~{\text{or       }}\forall_{n=1}^{N}\left(\left(\gamma_{\text{T}_n}+\gamma_{\text{F}}\right) < \gamma_{\text{th},N} ~\&~ \gamma_{\text{T}_{n-1}} \geq \gamma_{\text{th},(n-1)}\right)\Big]$ which ends up in the same result as in (\ref{S2SC}).
	
	%%%==============================================================
	\subsubsection{Diversity Order for the Combining Method}
	The diversity order of the $N$ cascaded THz links is  $\underset{\forall_{n=1}^{N}}{\min}\left(\frac{\xi_{\text{T}_n}^2}{2},\frac{\alpha_n N_{\text{r}}\mu_n}{2}\right)$. Further, the diversity order of the single FSO link is  $\min\left(\frac{\xi_{\text{F}}^2}{\kappa}, \frac{\alpha_{\text{F}}}{\kappa}, \frac{\beta_{\text{F}}}{\kappa}\right)$. Thus, the overall diversity order of System-2 for the combining method is obtained as
	\begin{align}\label{diersity_S2_ST}
		G^{\text{ST}}_{\text{d,S-2}}=\underset{\overset{\text{diversity order of the single E2E FSO link}}{}}{\underbrace{\min\left(\frac{\xi_{\text{F}}^2}{\kappa}, \frac{\alpha_{\text{F}}}{\kappa}, \frac{\beta_{\text{F}}}{\kappa}\right)}}+\underset{\overset{\text{diversity order of the N cascaded THz links}}{}}{\underbrace{\underset{\forall_{n=1}^{N}}{\min}\left(\frac{\xi_{\text{T}_n}^2}{2},\frac{\alpha_n N_{\text{r}}\mu_n}{2}\right)}}.
	\end{align}
	%%=================================================
	
	\subsection{Switching Method}
	
	\subsubsection{Outage Probability}	
	\begin{figure*}
		\begin{align}\label{S2Swi}
			\gamma^{\text{SW}}_{\text{CI}}=
			\begin{cases}
				\gamma_{\text{T}_{\text{eq}}}, & \text {for  } {\forall_{n=1}^N~\gamma_{\text{T}_n}} \geq \gamma_{\text{th},n}\\
				\gamma_{\text{F}}, & \text {for  } \forall_{n=1}^N\big(\gamma_{\text{T}_n} < \gamma_{\text{th},n} ~\&~ \gamma_{\text{F}}\geq \gamma_{\text{th}}~\&~ \gamma_{\text{T}_{n-1}}\geq \gamma_{\text{th},(n-1)}\big)\\
				0, & \text {for  } \forall_{n=1}^N\big(\gamma_{\text{T}_n} < \gamma_{\text{th},n} ~\&~ \gamma_{\text{F}}< \gamma_{\text{th}}~\&~  \gamma_{\text{T}_{n-1}}\geq \gamma_{\text{th},(n-1)}\big).
			\end{cases}
		\end{align}
		\hrulefill
	\end{figure*}
	
	\begin{table*}
		\centering 	
		\caption{Considered parameters for the simulations \cite{zedini2020performance,singyaperformance2022,li2022mixed,balti2019tractable}.}
		\begin{tabular}{|ll|ll|ll|}
			\hline
			\multicolumn{2}{|c|}{THz link}                                    &\multicolumn{2}{c|}{FSO link}                                    & \multicolumn{2}{c|}{mmWave access link}                               \\ \hline
			\multicolumn{1}{|l|}{Parameters}                      & Value     &	\multicolumn{1}{l|}{Parameters}                      & Value     & \multicolumn{1}{l|}{Parameters}                      & Value      \\ \hline
			\multicolumn{1}{|l|}{$f_{\text{T}_n}$}       & 119 GHz   & \multicolumn{1}{l|}{$\lambda_{\text{F}_n}$}       & 1550 nm & \multicolumn{1}{l|}{$f_\text{acc}$}       & 28 GHz     \\ 
			\multicolumn{1}{|l|}{$G_\text{t}^{\text{T}_n}$} & 55 dBi   &\multicolumn{1}{l|}{Strong: $C_n^2$} & 1$\times 10^{-12}$m$^{{-2}/{3}}$   & \multicolumn{1}{l|}{$G_\text{t}^\text{acc}$} & 44 dBi     \\ 
			\multicolumn{1}{|l|}{$G_\text{r}^{\text{T}_n}$}  & 55 dBi    & \multicolumn{1}{l|}{Moderate: $C_n^2$} & 5$\times 10^{-13}$m$^{{-2}/{3}}$    & \multicolumn{1}{l|}{$G_\text{r}^\text{acc}$} & 44 dBi     \\ 
			\multicolumn{1}{|l|}{$p$}      & 101325 Pa & \multicolumn{1}{l|}{Strong: $\alpha_{\text{F}_n}$,$\beta_{\text{F}_n}$}      & 4.343, 2.492 & \multicolumn{1}{l|}{$\varrho_{\text{ra}}$}             & 0 dB/km   \\ 
			\multicolumn{1}{|l|}{$T$}               & 298 K     & \multicolumn{1}{l|}{Moderate: $\alpha_{\text{F}_n}$,$\beta_{\text{F}_n}$ }               & 5.838, 4.249    &  \multicolumn{1}{l|}{$\varrho_{\text{ox}}$}   & 15.1 dB/km    \\ 
			\multicolumn{1}{|l|}{$\chi$}      & 50$\%$    & \multicolumn{1}{l|}{$\eta$}      & 1    & \multicolumn{1}{l|}{}                                &          \\ 
			\multicolumn{1}{|l|}{${a_{\text{T}_n}}$}             &  ${\lambda_{\text{T}_n}/{\left(2\pi\right)}\sqrt{G^{\text{T}_n}_{\text{t}}}}$    & 	\multicolumn{1}{l|}{${a_{\text{F}_n}}$}             & 20 cm    & \multicolumn{1}{l|}{}                                &            \\ 
			\multicolumn{1}{|l|}{${\omega_{\text{T}_n}}$}             & 50 cm     & \multicolumn{1}{l|}{${\omega_{\text{F}_n}}$}             & 40 cm    & \multicolumn{1}{l|}{}                                &            \\ 
			\multicolumn{1}{|l|}{$\varepsilon_{\text{T}_n}$}             &  6 cm    & 	\multicolumn{1}{l|}{$\varepsilon_{\text{F}_n}$}             & 5 cm    & \multicolumn{1}{l|}{}                                &            \\ \hline
		\end{tabular}
		\label{ParametersN}
	\end{table*}
	Since the FSO link is working as a backup for the THz link, first we check the availability of the THz link in the multi-hop hybrid THz/FSO-based backhaul network. 
	If all THz nodes can decode the data correctly, the equivalent multi-hop THz link will serve irrespective of the FSO link.  However, if any of the  THz links fails, the FSO link becomes active if $\gamma_{\text{F}} \geq \gamma_{\text{th}}$. 
	Therefore, for the considered System-2 with switching, the instantaneous received SNR at the final MiBS/child IAB is given as (\ref{S2Swi}), at the top of this page.
	From this, the outage probability is calculated as
	\begin{align}
		&\mathcal{P}^{\text{SW}}_{\gamma_{\text{CI}}}\left(\gamma_{\text{th}}\right)=\nonumber\\&
		\mathcal{P}\left[\forall_{n=1}^N\big(\gamma_{\text{T}_n} < \gamma_{\text{th},n} ~\&~ \gamma_{\text{F}}< \gamma_{\text{th}} ~\&~ \gamma_{\text{T}_{n-1}}\geq \gamma_{\text{th},(n-1)}\big)\right].\nonumber\\&
	\end{align}
	This can be solved as
	\begin{align}
		\mathcal{P}^{\text{SW}}_{\gamma_{\text{CI}}}\left(\gamma_{\text{th}}\right)&=F_{\gamma_{\text{F}}}\left(\gamma_{\text{th}}\right)\times\bigg[\sum_{n=1}^{N}F_{\gamma_{\text{T}_n}}\left(\gamma_{\text{th},n}\right) \nonumber\\&
		\times \prod_{n_1=1}^{n-1}\left(1-F_{\gamma_{\text{T}_{(n-n_1)}}}\left(\gamma_{\text{th},(n-n_1)}\right)\right)\bigg].
	\end{align}
	With non-IAB setup, the E2E outage probability of the considered network is derived as	
	\begin{align}\label{OUT_F_E2E_SC_S2}
		\mathcal{P}^{\text{SW}}_{\text{E2E,S-2}}\left(\gamma^{\text{UE}}_{\text{th}}\right)&=1-\left(1-\mathcal{P}^{\text{SW}}_{\gamma_{\text{CI}}}\right)\times\left(1-F_{\gamma_\text{acc}}\left(\gamma^{\text{UE}}_{\text{th}}\right)\right).
	\end{align}	
	Finally, the same as in (\ref{IAB_outage}), the outage probability in the out-band IAB setup is given by averaging the outage probability over all IAB donor and child IAB nodes.
	%%%%%%%%========================================
	\subsubsection{Diversity Order for the Switching Method}
	With DF relaying, the diversity order of the considered System-2 with switching is given as 
	\begin{align}
		G^{\text{SW}}_{\text{d,S-2}}=\underset{\overset{\text{diversity order of the single E2E FSO link}}{}}{\underbrace{\min\left(\frac{\xi_{\text{F}}^2}{\kappa}, \frac{\alpha_{\text{F}}}{\kappa}, \frac{\beta_{\text{F}}}{\kappa}\right)}}+\underset{\overset{\text{diversity order of the N cascaded THz links}}{}}{\underbrace{\underset{\forall_{n=1}^{N}}{\min}\left(\frac{\xi_{\text{T}_n}^2}{2},\frac{\alpha_n N_{\text{r}}\mu_n}{2}\right)}}.
	\end{align}
	Note that the diversity order of System-2 for both combining and switching methods is the same, which is already proved in Section III for System-1.
	%%%%%%%%========================================
	
	\section{Mesh Network}
	We consider a mesh network with $Q$ independent and non-overlapping multi-hop routes from the MaBS to the UE. The $q^{\text{th}}$ route ($q=1,2,...,Q$) consists of the multi-hop hybrid THz/FSO links according to System-1 and System-2.  
	Thus, for the mesh network, outage probability is
	\begin{align}\label{Mesh_out}
		\mathcal{P}^{\text{mesh}}_{\text{out}}=\prod_{q=1}^{Q}\mathcal{P}_{\text{out},q},
	\end{align}
	where $\mathcal{P}_{\text{out},q}$ is the outage probability of the $q^{\text{th}}$ E2E path from the MaBS to the destination UE. For System-1, this is derived in (\ref{OUT_F_E2E_MRC}) and (\ref{OUT_F_E2E_S1}) for the combining and the switching methods, respectively.
	For System-2, this is derived in (\ref{OUT_F_E2E_MRC_S2}) and (\ref{OUT_F_E2E_SC_S2}) for the combining and the switching methods, respectively. Note that in (\ref{Mesh_out}) we have
	used the fact that in a mesh network an outage occurs if the data is correctly transferred to the destination through none of the routes  and there is no overlap between the routes.
	
	%%%%%%%%========================================
	\section{Simulation Results}
	We present our results with respect to the transmit power which is defined as $10\log_{10}\left({P_{\text{j}}}\right)$ in the log-domain. Here, $P_{\text{j}}$ with $\text{j}\in\{\text{T,F,acc}\}$ is the transmit power at the $j$ node. For simplicity, we consider $P_{\text{F}}=P_{\text{T}}=P_{\text{acc}}$ in each hop for the analysis. Then, $\sigma^2_{\text{o}}=\sigma^2_{\text{T}}=\sigma^2_{\text{acc}}=1$ is assumed in each hop. {The infinite summation present in eq. (28) is truncated to 50 values for improved accuracy with acceptable computational complexity.} {Further, $10^8$ iterations are performed for Monte-Carlo simulations.} Rest of the simulation parameters are given in Table \ref{ParametersN}. {Note that we mention all the practical parameters related to the THz, FSO, and mmWave links in Table \ref{ParametersN}. Such parameters are also preferred in \cite{zedini2020performance,singyaperformance2022,li2022mixed,balti2019tractable}}.
	
\begin{figure}[t!]
		\centering
		\includegraphics[width=3.5in,height=2.6in]{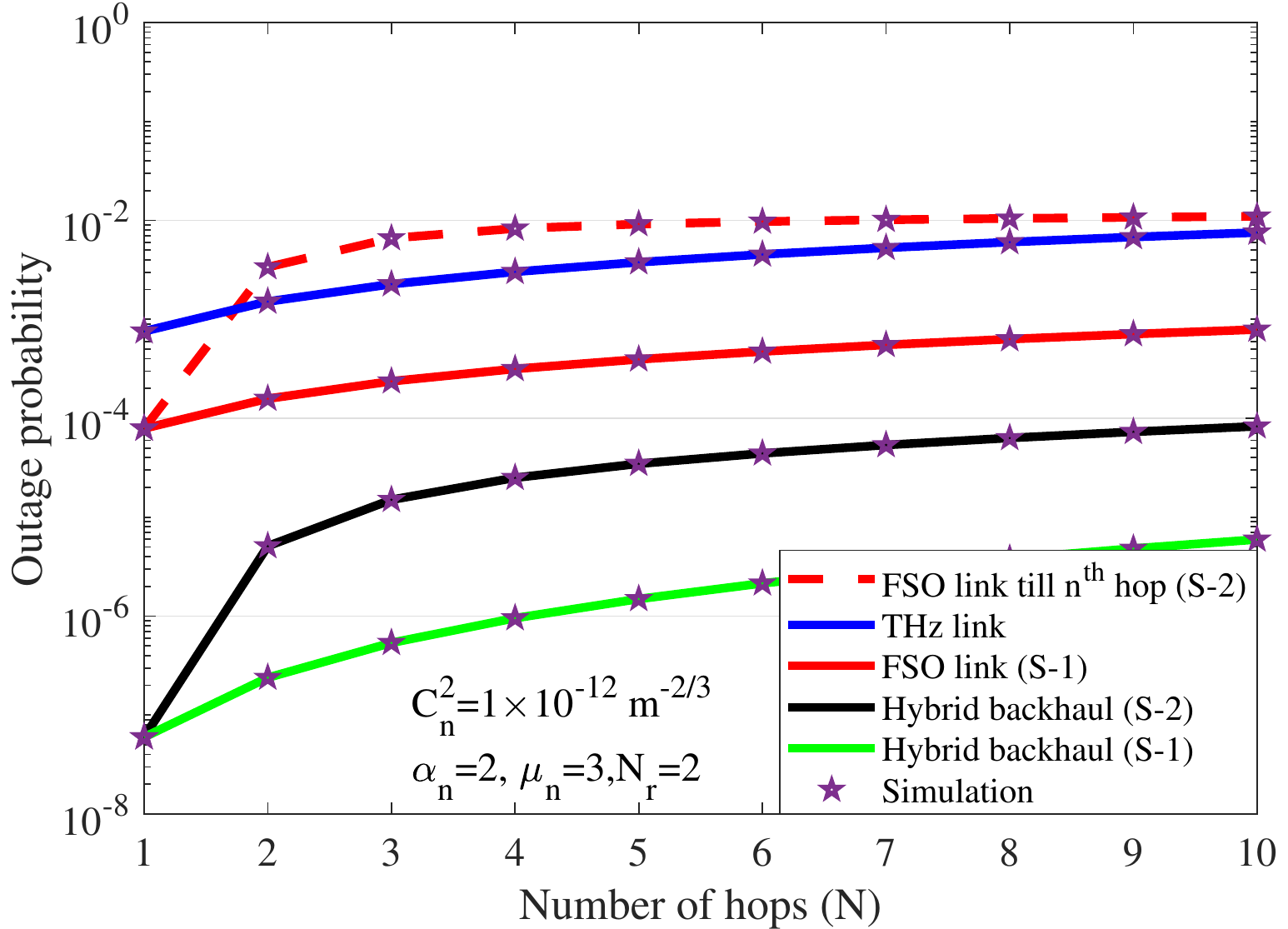}
		\caption{\small{Impact of number of hops on the outage probability of  the backhaul hybrid THz/FSO systems with switching. Here, we consider 200 m links in each hop, 25 dB transmit power, $\gamma_{\text{th}}=1$ dB in each hop,  and present the results for both System 1 and 2. Also, $C_n^2=1\times 10^{-12} \text{m}^{-2/3}$ and $\alpha_{n}=2,\mu_{n}=3,N_r=2$.}}
		\label{NHopsS1}
	\end{figure}

\begin{figure}[h!]
		\centering
		\includegraphics[width=3.5in,height=2.6in]{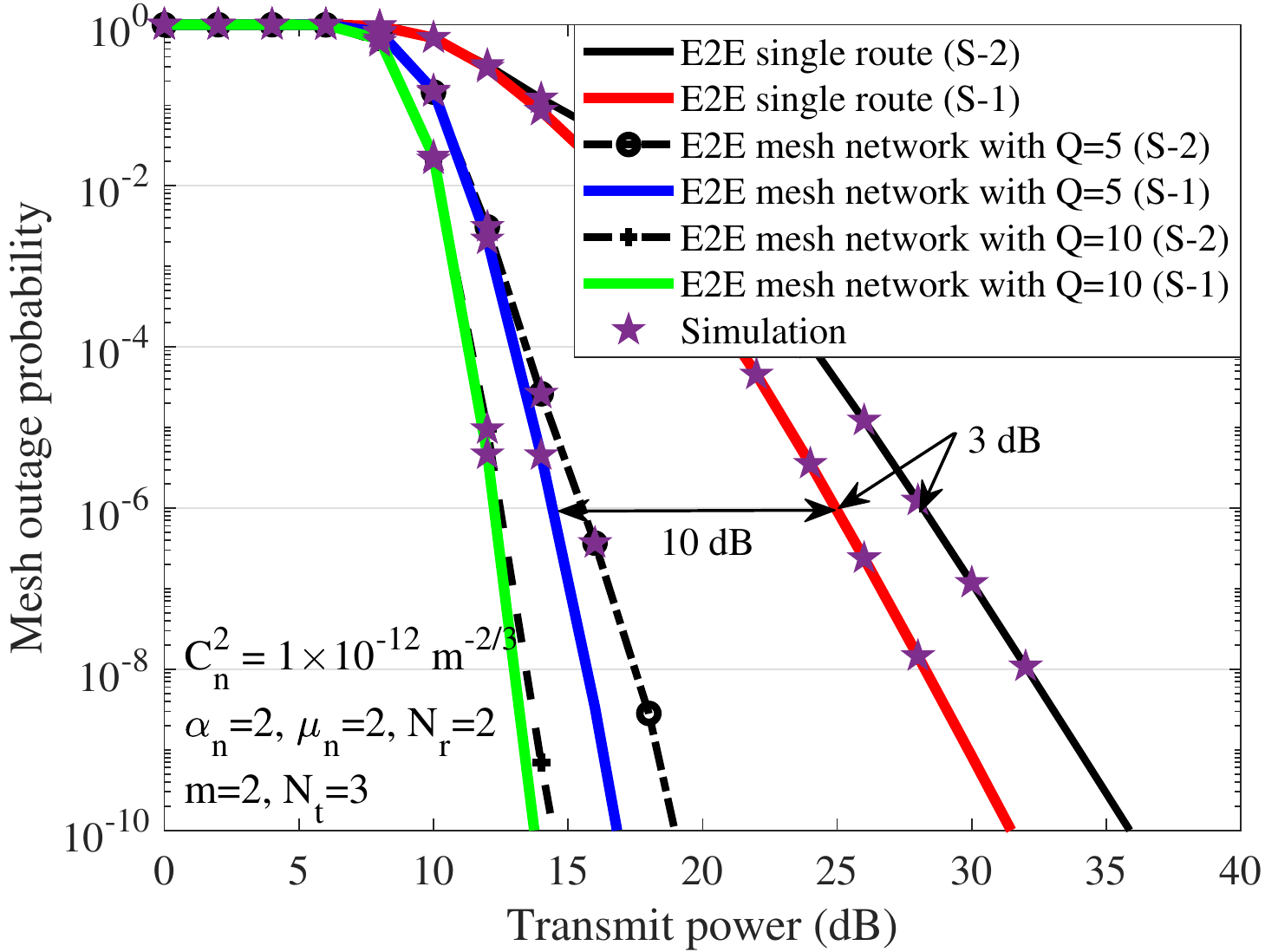}
		\caption{\small{E2E outage probability of mesh network for both the systems with different routes by fixing 3 hops per route (2 backhaul and one access link) and by considering switching method. Here, $L_{\text{F}}=200$ m, $L_{\text{T}}=200$ m, and $L_{\text{acc}}=100$ m in each hop.}}
		\label{Mesh}
	\end{figure}

\begin{figure}[b!]
		\centering
		\includegraphics[width=3.5in,height=2.7in]{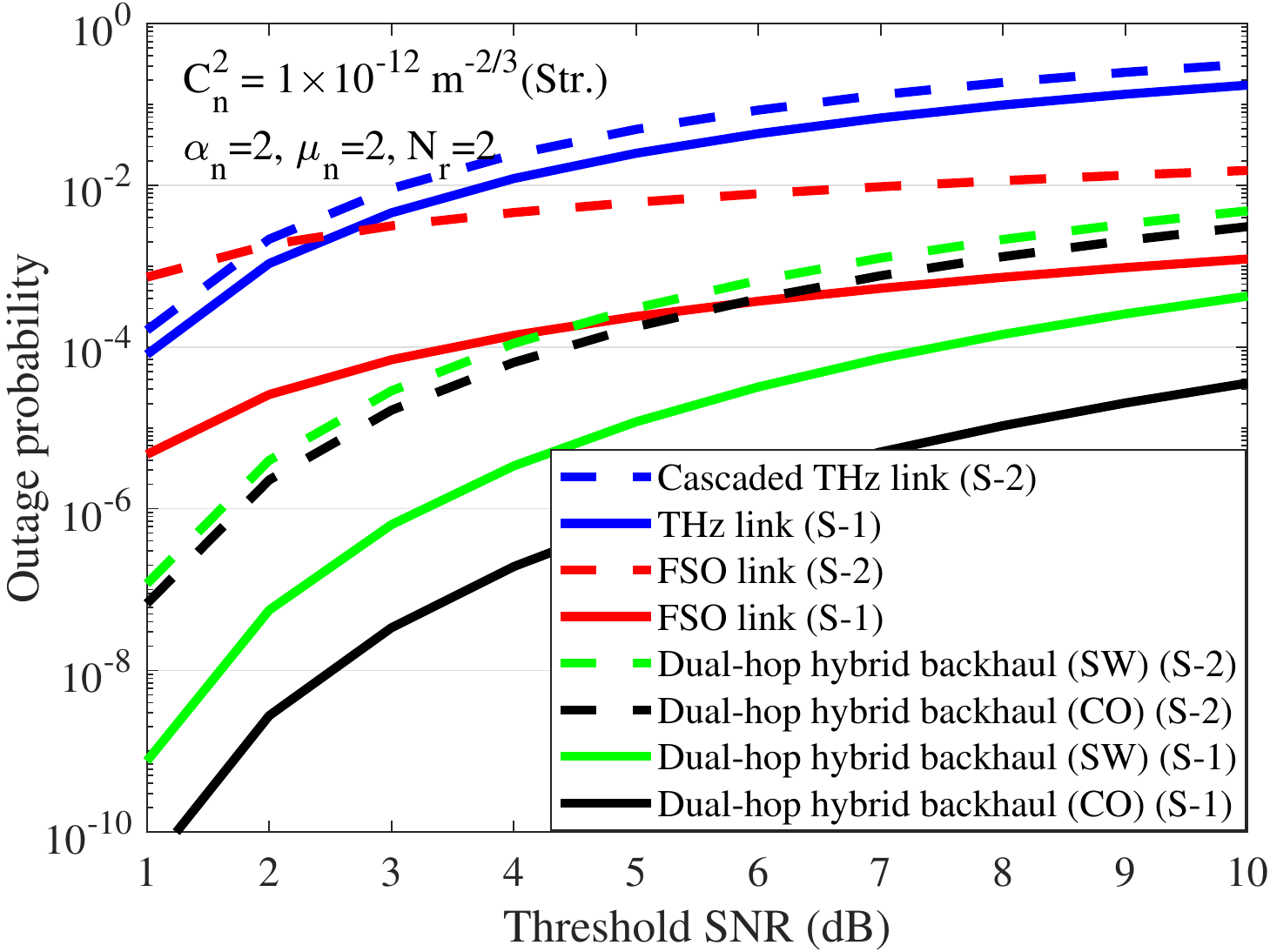}
		\caption{\small{Comparison of outage probability of both the multi-hop THz/FSO-based systems against SNR threshold.}}
		\label{OutS1S2_thr}
	\end{figure}

\begin{figure*}[hb]
		\centering
		\includegraphics[width=7.2in,height=4.5in]{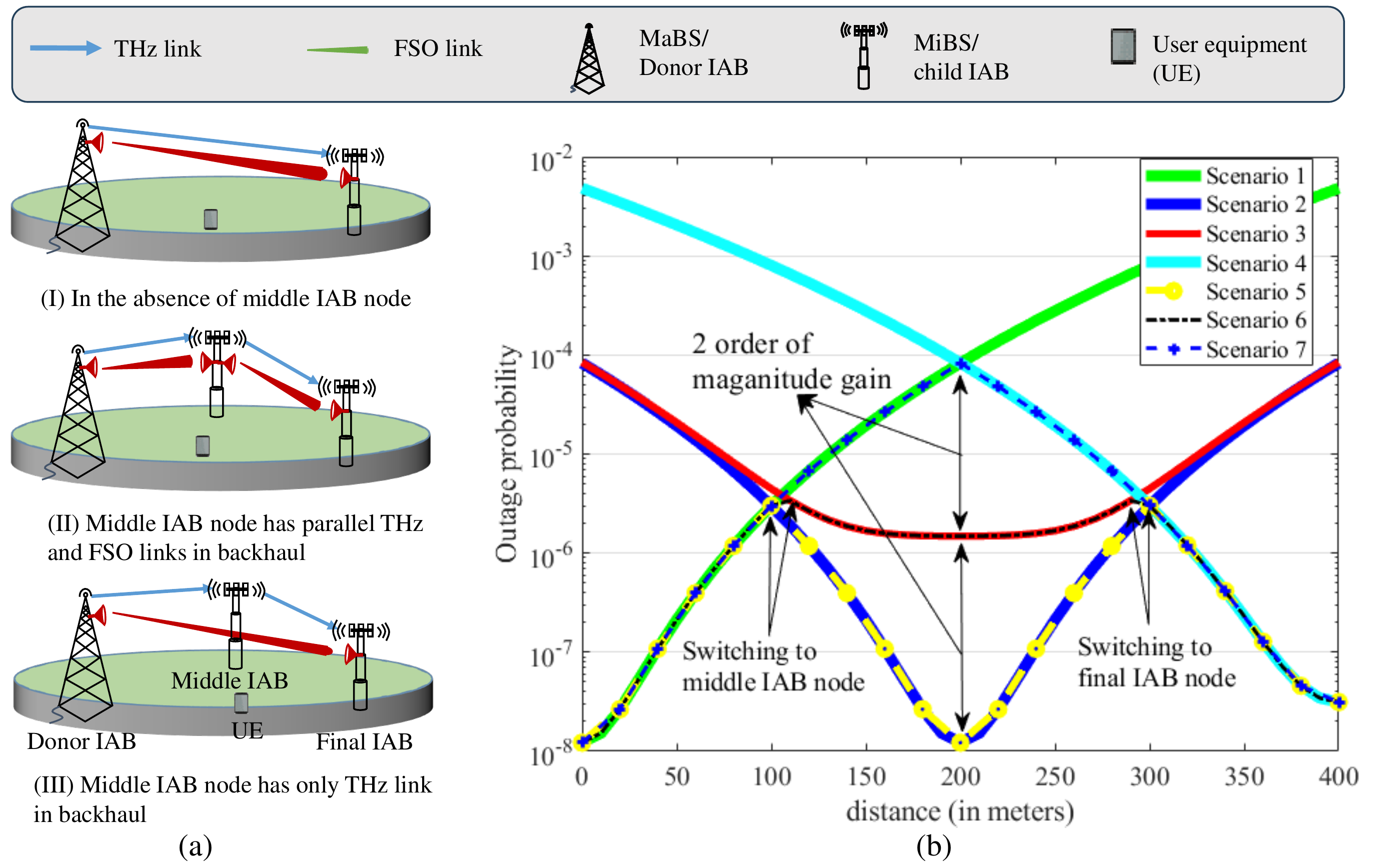}
		\caption{{(a) Various network configurations. (b) Outage probability at the MU while moving from donor-middle IAB-final IAB node. }}
		\label{Out_Main}
	\end{figure*}

	  We present the results for the general cases with multi-hop and mesh networks. However, as demonstrated in \cite{madapatha2020integrated,madapatha2021topology}, with multi-hop and IAB networks, the system performance is significantly affected by increasing the number of hops, as the traffic congestion and E2E latency increases. For this reason, except for Fig. \ref{NHopsS1}, where multi-hop results are shown for both the systems, only dual backhaul hops are considered for rest of the results. In Figs. \ref{Out_Main}-\ref{AsymS1S2}, we present the results for the cases with IAB setup where each intermediate IAB node in the multi-hop setup serves a number of surrounding UEs. In Figs. \ref{NHopsS1}-\ref{OutS1S2_thr}, on the other hand, we present the results for the non-IAB setup where the UEs are served only by the end MiBS. Note that if the SNR threshold is fixed in all hops, we are in non-IAB case (even if the figure only presents the backhaul performance).
	
	%%============================================		
	
	In Fig. \ref{NHopsS1}, we compare the outage probability of the multi-hop backhaul hybrid THz/FSO systems for different turbulence/fading conditions. We study both System-1 (S-1 in Figs.) and System-2 (S-2 in Figs.), as defined in Figs. \ref{System1_NLoS}-\ref{System2_LoS}, respectively.
	From Fig. \ref{NHopsS1}, we observe that the FSO link's performance decreases significantly for System-2 (shown with dotted red) as compared to System-1 (shown with solid red), where $N$ FSO links are deployed in DF mode. However, for both systems, $N$ THz links are deployed in series for the E2E backhaul communication, hence, the THz link's performance remains the same (shown in blue for both systems). Consequently, the E2E performance of System-2 decreases significantly as compared to System-1. As shown in Fig. \ref{NHopsS1}, for both systems, increasing the number of hops from 1 to 2 results in significant outage probability increment. However, for more than 2 hops, the outage probability is only slightly affected by increasing the number of hops. On the other hand, as discussed before and shown in \cite{madapatha2020integrated,madapatha2021topology}, the E2E delay/throughout are significantly deteriorated by increasing the number of hops.

	In Fig. \ref{Mesh}, we show the E2E outage probability of mesh networks for both system models. Here, in each route, we fix the number of hops to three with two hybrid THz/FSO-based backhaul hops and one mmWave access hop. We consider strong turbulence/fading case by setting $C_n^2=1\times 10^{-12}$ m$^{-2/3}$ which gives $\alpha_{\text{F}}=4.34$, $\beta_{\text{F}}=2.49$ for the FSO links. Further, we consider $\alpha=2, \mu=2, N_{\text{r}}=2$ and $m=2, N_{\text{t}}=3$ for the THz and mmWave access links, respectively. \par
	
	As compared to the single route multi-hop network (as shown in Fig. \ref{NHopsS1}), whose outage performance degrades with number of hops, a mesh network improves the overall outage performance by enabling more routes. 
	From Fig. \ref{Mesh}, we observe that for the considered set of parameters  and an outage probability $10^{-6}$, System-1 provides 3 dB gain over System-2 for a single route multi-hop network. Also, by increasing the E2E routes ($Q=5$) in the mesh network  and outage probability $10^{-6}$, we obtain approximately 10 dB gain over the single route multi-hop network for both systems. Adding more routes will further improve the E2E performance, however, with a decreased rate than before. The results are shown for the switching case. With combining, we will achieve improved performance over switching, however, the qualitative insights remain the same.	\par

	In Fig. \ref{OutS1S2_thr}, we show the impact of  SNR threshold on the outage performance of the dual-hop backhaul systems for both System-1 (solid lines) and System-2 (dotted lines) with both
	switching and combining methods. Here, we consider the same SNR threshold at each receiving MiBS node, hence, it is a non-IAB system. The FSO link of System-1 outperforms the FSO link of System-2, since, for the dual-backhaul hops, the FSO link in System-2 is approximately twice longer than the FSO link in System-1. Further, the cascaded THz link of System-2 (dual-hop) has slightly lower performance than the single THz link of System-1. Consequently, the dual-hop backhaul hybrid THz/FSO-based link of System-1 has better performance than System-2 for both the switching and combining methods. Further, we observe a significant performance improvement in System-1 with combining over the switching method at the receiving MiBS as compared to System-2. This is due to the fact, that in System-2, the switching/combining is performed at the end of the second hop, and the single THz link of the first hop dominates in the backhaul dual-hop communication. Hence, in this case, we observe only slight improvement in combining over switching in System-2.

	In Figs. \ref{NHopsS1}-\ref{OutS1S2_thr}, we have shown the results of  non-IAB multi-hop and mesh networks, where identical SNR threshold is considered on each of the receiving MiABs.
	On the other hand, Figs. \ref{Out_Main}-\ref{AsymS1S2} illustrates the results for IAB networks. With IAB, each IAB node carries the information of all UEs to be served by the IAB node itself and its following child IAB nodes. Different SNR thresholds consideration at each IAB node is motivated with the fact that the SNR threshold is directly related to the number of UEs associated with each IAB node. Let us assume a desired rate $\mathcal{R}$ bps/Hz at each UE, and $L$ such UEs are being served by the $n^{\text{th}}$  IAB node or its following child IAB nodes in the multi-hop chain. Then, the SNR threshold at the  $n^{\text{th}}$  IAB node will be $\gamma_{\text{th},n}=2^{L\times \mathcal{R}}-1$. Thus, the SNR threshold of each IAB node is scaled by the total number of UEs which are served by the IAB node and its following child nodes.\par
	%	Let, $L_{\text{I}_0}$, $L_{\text{I}_1}$, and $L_{\text{I}_2}$ are the number of UEs being served by the MaBS/donor IAB, first MiBS/child IAB, and the second MiBS/child IAB, respectively. Then the SNR threshold at the second IAB will be $2^{L_{\text{I}_2}\times \mathcal{R}}-1$, at the first MiBS/child IAB will be $2^{\left(L_{\text{I}_1}+L_{\text{I}_2}\right)\times \mathcal{R}}-1$, and at the MaBS/donor IAB will be $2^{\left(L_{\text{I}_0}+L_{\text{I}_1}+L_{\text{I}_2}\right)\times \mathcal{R}}-1$.\par

{Considering 10 UEs associated to each IAB node with $\mathcal{R}=0.1$ bps/Hz, we set the SNR threshold at each IAB node. In Fig. \ref{Out_Main}, we analyze the outage performance of the UE while moving from the donor to middle IAB and then to the last child IAB node. 
	We consider that the height of the donor and the child IAB node is 30 m and the length of each hop is 200 m. We consider following scenarios to serve a UE while moving from donor to middle and then to the final child IAB node:
	\begin{itemize}
		\item Scenario 1: UE is always directly connected to donor IAB node. 
		\item Scenario 2: UE is always connected through the middle IAB node via access link. The backhaul link from donor to middle IAB node is parallel THz-FSO link.
		\item Scenario 3: Same as Scenario 2, but the backhaul link from donor to middle IAB node is THz link only. 
		\item Scenario 4: UE is always connected through the final child IAB node. The backhaul link is a 400 m long parallel THz-FSO link in the absence of middle IAB node.
		\item Scenario 5: A dual-hop donor-middle IAB-final child IAB network is serving the UE via access link. The dual-hop backhaul link is having parallel deployment of the THz and the FSO links at each node. 
		\item Scenario 6: Same as scenario 5. The only difference is that the dual-hop backhaul has THz link only between the donor and middle child IAB node. FSO transceiver is installed at donor and final child IAB node only.
		\item Scenario 7: Single hop 400 m long backhaul donor-final child IAB network is serving the UE via access link, i.e., the middle node is not considered.. 
		\end{itemize}
	In Fig. \ref{Out_Main}, we illustrate the outage performance of the UE in different scenarios while moving in-between different IAB nodes.
	In Scenario 1, when UE is directly connect to the donor IAB, the outage probability increases significantly with the movement of UE away from the donor IAB. The outage probability reaches till $10^{-2}$ when UE reaches to 400 m away from the donor IAB node. Similar performance is seen in Scenario 4 while UE is going towards the final IAB node which is serving the UE via 400 m parallel THz-FSO link followed by the access link. Note that, in both Scenario 1 and 4, middle IAB node is absent.
	 Therefore, considering Scenario 7 where UE is served by the 400 m long backhaul donor-final IAB network, outage probability at UE is maximum when it is in between the donor and final IAB node. Hence, to guarantee the best optimum coverage performance at the UE,  short backhaul links with intermediate nodes are necessary. \par
	 
	 	\begin{figure}[b]
	 	\centering
	 	\includegraphics[width=3.4in,height=2.7in]{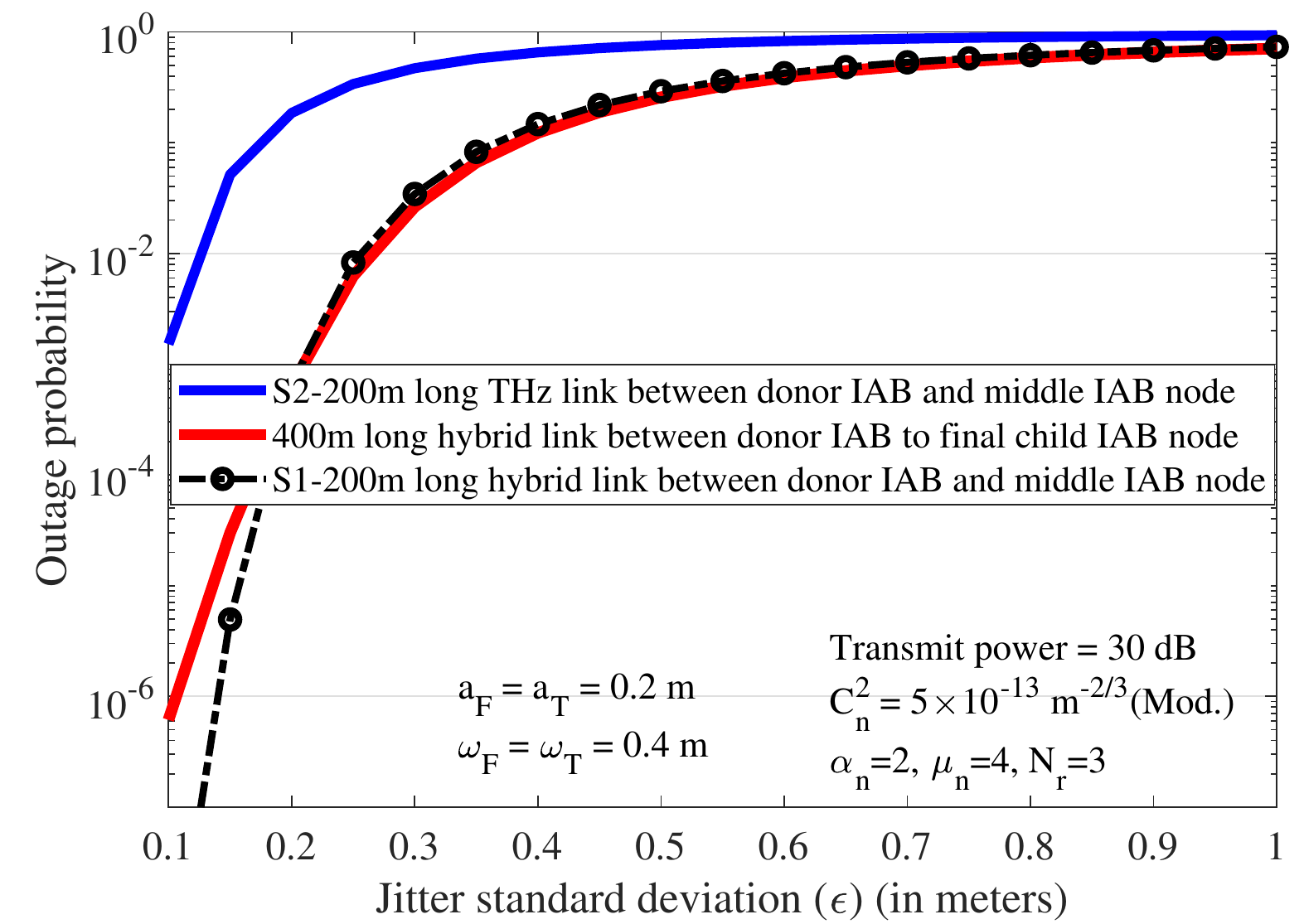}
	 	\caption{{Outage probability versus jitter standard deviation.}}
	 	\label{point_Main}
	 \end{figure}

	 \begin{figure}[]
	 	\centering
	 	\includegraphics[width=3.5in,height=2.6in]{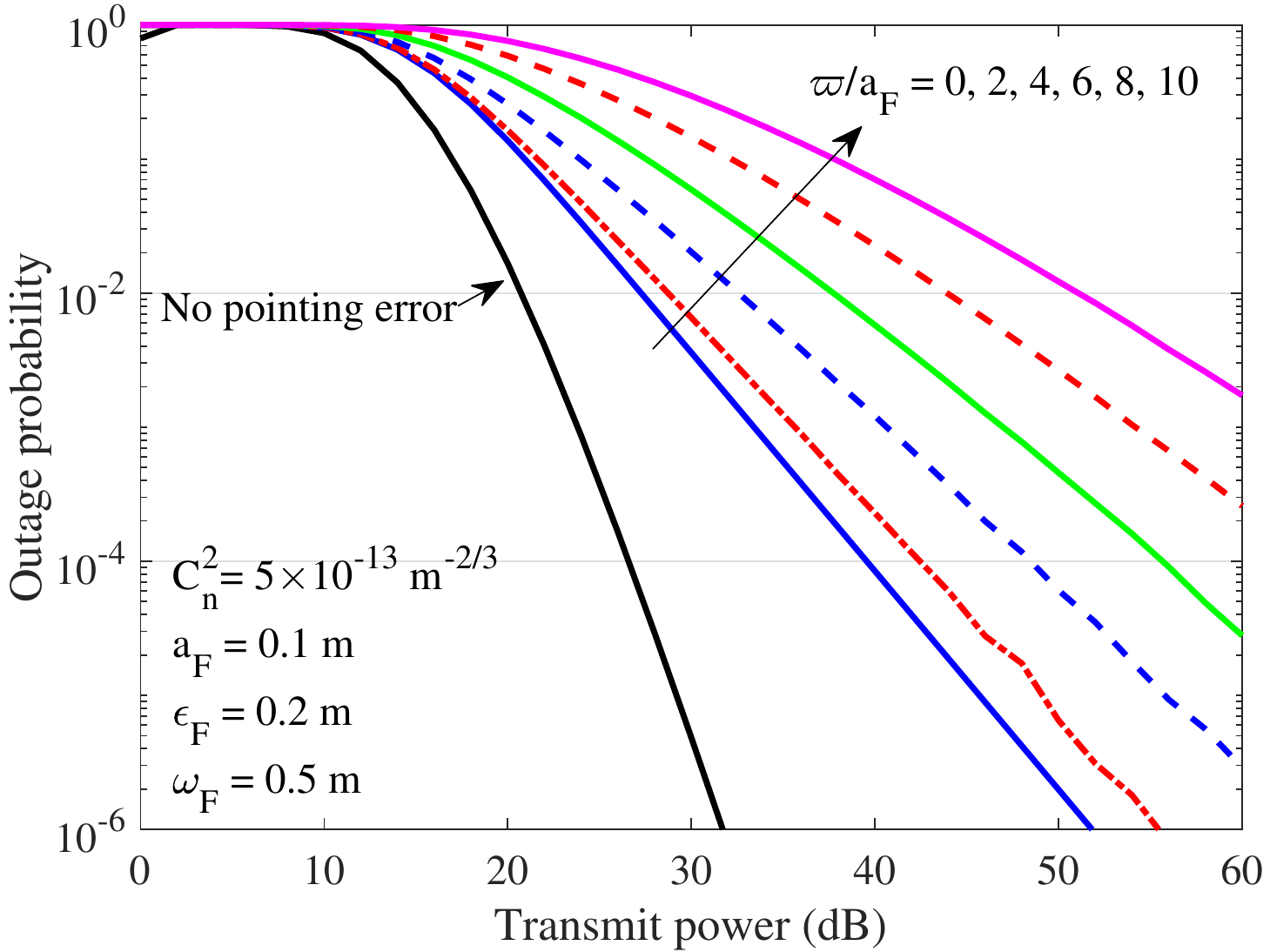}
	 	\caption{\small{Simulation results for the outage probability of FSO link against transmit power. For the results, we consider no pointing error, pointing error with only jitter variance, and pointing error with combined impact of boresight error and jitter variance cases.  }}
	 	\label{pnt_bor}
	 \end{figure}

	 \begin{figure*}[h]
	 	\hfill
	 	\centering
	 	\subfloat[\small{Strong turbulence/fading condition}]{\includegraphics[width=8.2cm,height=6.6cm]{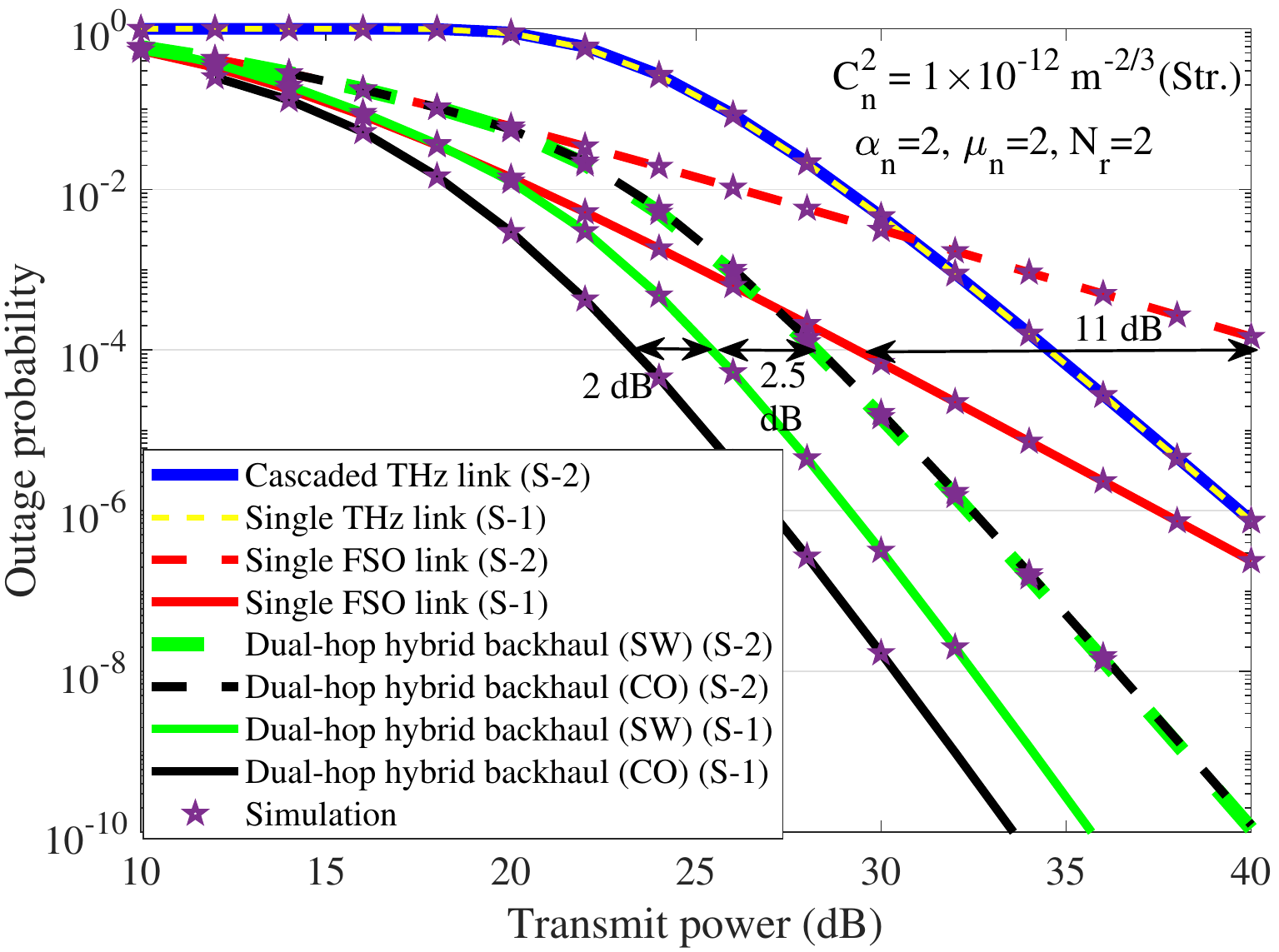}
	 		\label{S1S2_str}}
	 	\hfill
	 	\subfloat[\small{Moderate turbulence/fading condition}]{\includegraphics[width=8.2cm,height=6.6cm]{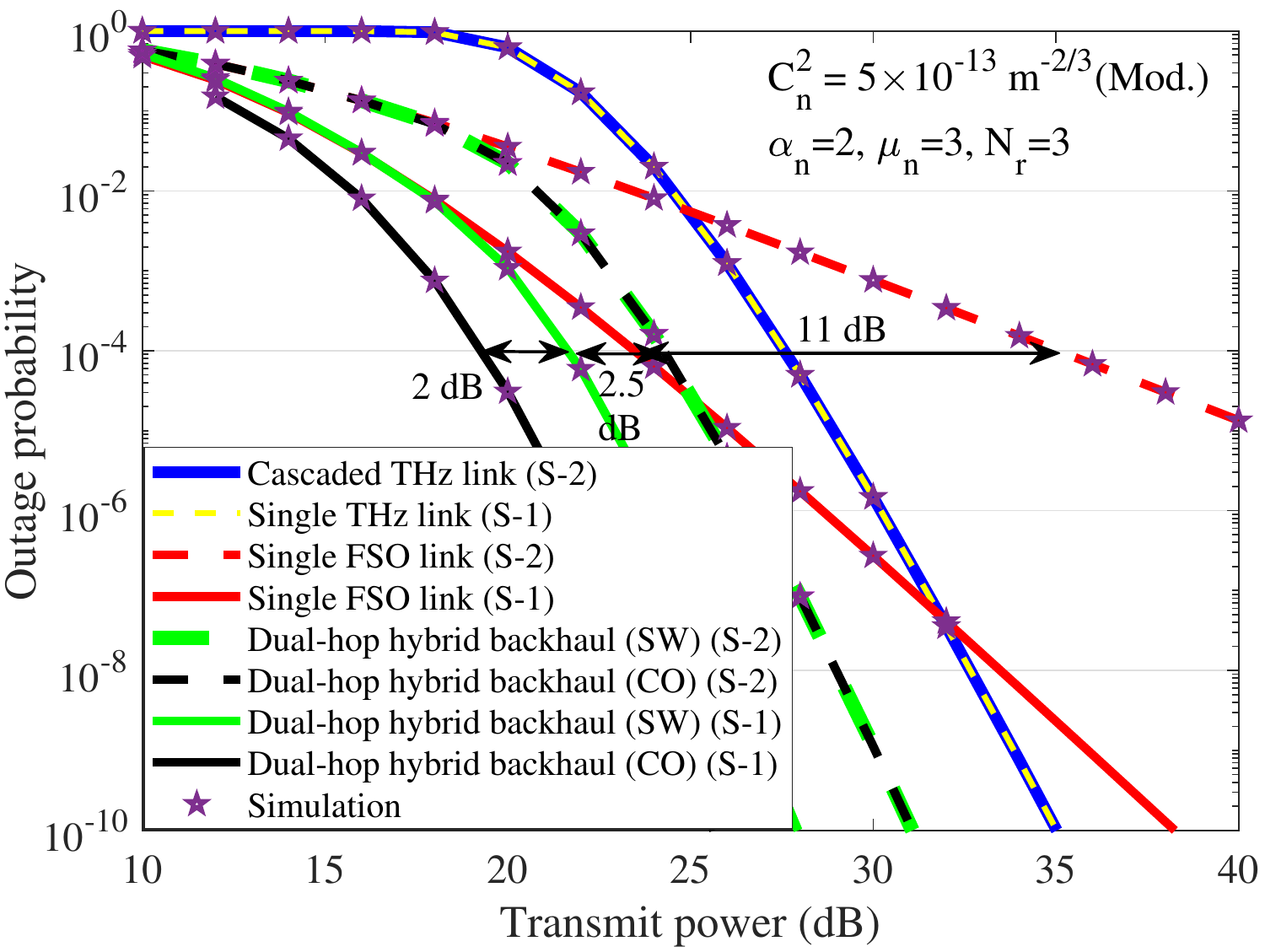}
	 		\label{S1S2_mod}}
	 	\hfill
	 	\caption{\small{Comparison of outage probability for both system models versus transmit power.}}
	 	\label{OutS1S2}	
	 \end{figure*}
	 
	 Considering Scenarios 2 and 3, where middle IAB node is always serving the UE, best outage performance is observed when UE is near to the middle IAB node. Outage performance decreases when UE moves away from the middle IAB node towards donor or final IAB node. Further, outage performance at UE is superior in Scenario 2 than in Scenario 3 by 2 orders of magnitude, because in Scenario 2, middle IAB node is assisted with the parallel deployment of the THz and FSO links in backhaul, as opposed to Scenario 3 where only the THz link is available in backhaul.
	 
	  Considering Scenarios 5 and 6, with the UE's movement, it is first serviced by the donor node, then connected to the middle IAB node, and finally switched to the final IAB node. 	 
	 Thus, considering the whole coverage area, in Scenarios 5 and 6, outage performance remains more uniform than in Scenario 1 and 4. Conclusively, outage performance at the UE always remains in acceptable range.
	For	example, in Scenarios 5 and 6, at 200 m,  we achieve nearly 4 and 2 orders of magnitude gain in outage performance respectively, as opposed to Scenario 4. Further, Scenario 5 provides improved outage performance than Scenario 6 for all ranges due to the availability of parallel THz and FSO links in backhaul. Finally, while we have not considered the soft-handover in the simulations, the benefits of the middle IAB node becomes even more visible in the cases with soft-handover. \par

	{In Fig. \ref{point_Main}, we show the impact of pointing error by varying the jitter standard deviation ($\varepsilon_{\text{F}}=\varepsilon_{\text{T}}=\varepsilon$) on the performance of the THz, the FSO, and the hybrid THz/FSO links. For this, we fix the receiver radius $a_{\text{F}}=a_{\text{T}}=0.2$ m and beamwidth $\omega_{\text{F}}=\omega_{\text{T}}=0.4$ m in each hop. 
	From Fig. \ref{point_Main}, we observe that the 200 m hybrid THz/FSO link between the donor and middle IAB node (System-1) has acceptable performance till $\varepsilon=0.25$ m (for an outage probability of $10^{-2})$, and reaches into outage with further increase in $\varepsilon$. The 400 m hybrid THz/FSO link between the donor and final IAB node (when middle IAB is not used) has poorer performance than the 200 m hybrid THz/FSO link at lower $\varepsilon$, however, has nearly the same performance at higher values of $\varepsilon$. Also, the 400 m hybrid THz/FSO link has acceptable performance till $\varepsilon=0.25$ m (for an outage probability of $10^{-2}$), and reaches into outage with further increase in $\varepsilon$. 
	On the other hand, 200 m THz link between the donor and middle IAB node (System-2) has the worst performance, and outage probability goes larger than $10^{-2}$  even at approximately $\varepsilon=0.15$ m. Further, increase in $\varepsilon$ forces the THz link into complete outage. Hence, from Fig. \ref{point_Main}, at low values of $\varepsilon$, System-1 is more sensitive to pointing error, which is intuitively because, in System-1, in each hop both the FSO and the THz links are affected by the pointing error.}

	{  Further, Fig. \ref{pnt_bor} illustrates the impact of pointing error in the FSO link by considering the combined impact of boresight error and jitter variance. For  this, we consider nonzero boresight error by setting nonzero values of $\varpi_{\text{x,j}_n}$ and $\varpi_{\text{y,j}_n}$ with $\varpi=\sqrt{\varpi^2_{\text{x,j}_n}+\varpi^2_{\text{y,j}_n}}$ and $\varepsilon_{\text{x,j}_n}^2=\varepsilon_{\text{y,j}_n}^2=\varepsilon_{\text{j}_n}^2$ for the $n^{\text{th}}$-hop.  For simulation, we consider the moderate turbulence case by setting  $C_n^2=1\times 10^{-12} \text{m}^{-2/3}$. Further, we consider the  receiver radius, beamwidth, and jitter standard deviation of the FSO link  $a_\text{F} = 0.1 $ m, $\omega_{\text{F}} = 0.5$ m, and $\epsilon_\text{F} = 0.2 $ m, respectively. From Fig. \ref{pnt_bor}, we observe that without pointing error, the FSO link has far superior performance than the cases with pointing error. Specifically, for an outage probability of $10^{-4}$, FSO link provides nearly 7 dB gain in no pointing error case as compared to the case with jitter variance and no boresight error induced pointing error. The performance of FSO link further degrades with the existence of boresight error along with the jitter variance, as shown in Fig. \ref{pnt_bor}.}

	%%%================================= 
	
{	Considering 10 UEs associated to each IAB node with $\mathcal{R}=0.1$ bps/Hz, we set the threshold at each IAB and obtain the outage results for both systems with switching/combining methods in Fig. \ref{OutS1S2}. Here, Fig. \ref{S1S2_str} and Fig. \ref{S1S2_mod} show the results for the strong and moderate turbulence/fading, respectively.
	From Fig. \ref{S1S2_str}, we observe that the FSO link of System-1 has significant performance improvement (around 11 dB for an outage probability $10^{-4}$) than the FSO link of System-2. This allows superior outage performance for System-1, as compared to System-2 for both switching/combining methods (around 2.5 dB gain with switching and 4.5 dB gain with combining method for an outage probability $10^{-4}$). Also,  the combining method provides around 2 dB gain over the switching method for System-1. However, the outage performance remains the same for both switching/combining methods for System-2. This is because the switching/combining is performed at the second hop and the single THz link of the first hop dominates in overall performance. The similar insights can be obtained from Fig. \ref{S1S2_mod} for the moderate turbulence/fading case.}			
	\begin{figure}[h]
		\centering
		\includegraphics[width=3.5in,height=2.6in]{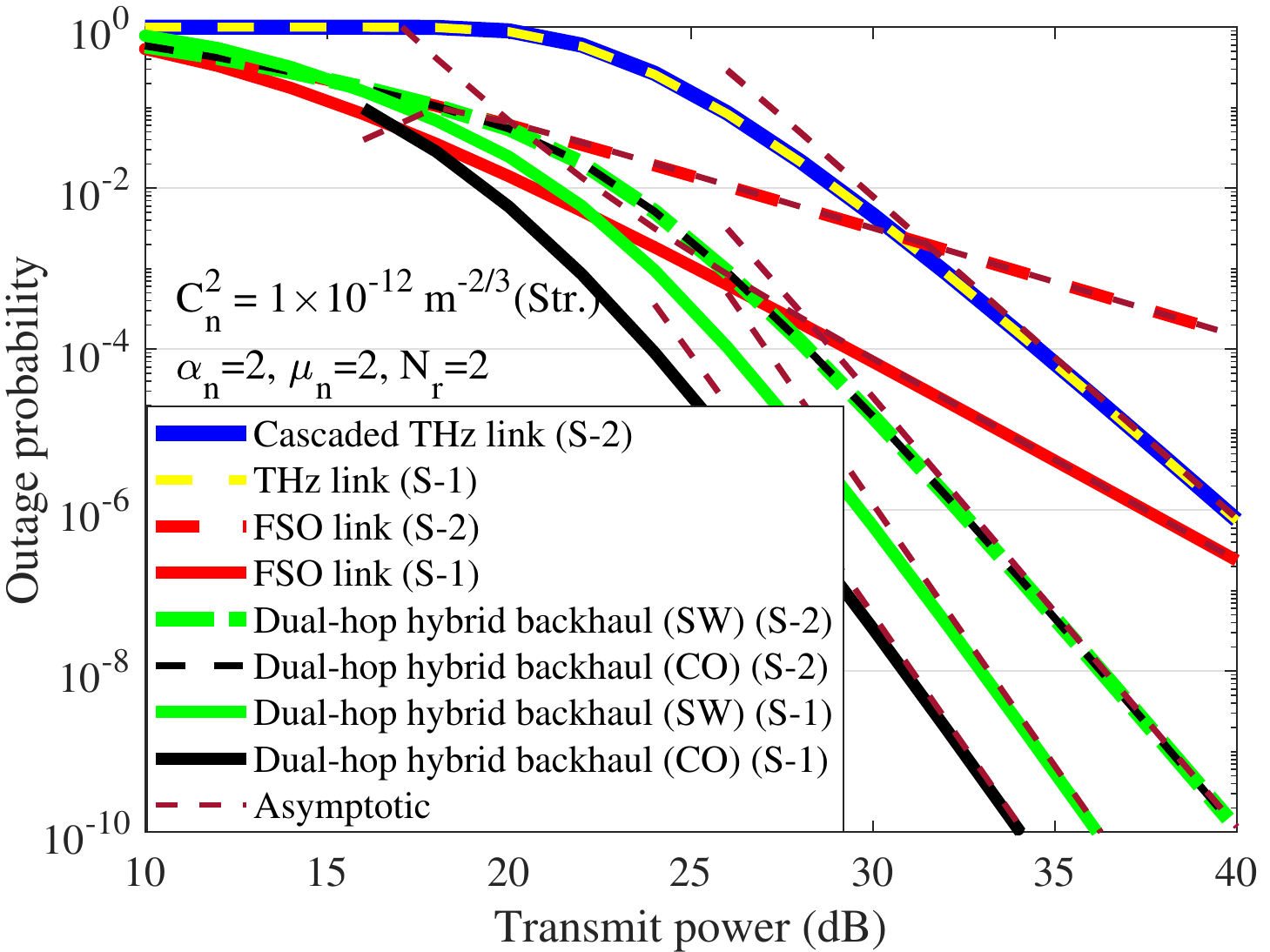}
		\caption{\small{Theoretical and asymptotic outage probability of both the systems for combining/switching methods.}}
		\label{AsymS1S2}
	\end{figure}
	
	In Fig. \ref{AsymS1S2}, the asymptotic outage probability results are shown for both systems. The asymptotic outage results follow the exact outage probability at medium and high SNRs which validates their accuracy. Further, at high SNRs, the results of combining and switching have the same slope for both systems which justifies the same diversity order for the combining and switching methods. The diversity order of the FSO link is  $\text{min}\left(\frac{\xi_{\text{F}_n}^2}{\kappa}, \frac{\alpha_{\text{F}_n}}{\kappa}, \frac{\beta_{\text{F}_n}}{\kappa}\right)$ and the diversity order of the THz link is $\min\left(\frac{\xi_{\text{T}_n}^2}{2},\frac{\alpha_n N_{\text{r}}\mu_n}{2}\right)$. Thus, with multi-hop parallel THz/FSO deployment, the diversity order of System-1 for both combining and switching is given by (\ref{diversity_S1_SW}).
	Similarly, the diversity order of System-2 for both the combining/switching methods is obtained by (\ref{diersity_S2_ST}).	
	%%======================================================================
	\section{Conclusions}
	In this work, we consider multi-hop and mesh hybrid THz/FSO-based backhaul networks with different deployments of the THz and the FSO links, and analyze their performance for the cases with both IAB and non-IAB based communication setups. For the analysis, we consider both the combining and switching methods between the THz/FSO links, and the impact of various practical considerations like atmospheric attenuation/path-loss, pointing/misalignment error, atmospheric turbulence/fading, number of antennas, number of UEs, number of hops, and the threshold data-rates are shown on the performance of considered systems.
	From the results, we observe that the parallel deployment of the THz/FSO links in the backhaul with proper system configuration improves the overall performance of the communication system. Also, the diversity order for combining and switching methods remains the same. However, combining provides a considerable performance gain, compared to switching. Further, the increased jitter standard deviation drastically reduces the performance of the THz, the FSO, and consequently the backhaul THz/FSO link. This significantly deteriorates the E2E performance.  Then, an increase in number of hops increases the coverage area, at the cost of traffic congestion and E2E latency. Also, mesh networks enable multiple routes and improve the E2E performance significantly.  {Finally, it is interesting to note that practical implementation challenges of THz/FSO link include, e.g., synchronization between the THz and FSO  links, security aspects (where the FSO link is more secure than a THz link) as well as deployment issues because two different types of transceivers need to be deployed at the same point, increasing the deployment cost.}
	%%%=================================
	
%	\section{Acknowledgment} We would like to thank Prof. M.-S. Alouini and KAUST, Saudi Arabia for the support.
	\appendices
	\numberwithin{equation}{section}
	\section{Proof of Outage Probability for Combining Method}
	
	The (\ref{outMRCSC}) can be solved as
	{\begin{align}\label{outMRC2}
			&\mathcal{P}^{\text{ST}}_{\gamma_{\text{I}_n}}(\gamma_{\text{th},n})=\int_{0}^{\gamma_{\text{th},n}} f_{\gamma_{\text{F}_n}}(\gamma)F_{\gamma_{\text{T}_n}}(\gamma_{\text{th},n}-\gamma) \text{d}\gamma, \nonumber\\&=
			\frac{\xi^2_{\text{F}_n}}{\kappa\Gamma(\alpha_{\text{F}_n})\Gamma(\beta_{\text{F}_n})}\frac{\mathbb{C}^n_{1}}{\xi_{\text{T}_n}^2}
			\nonumber\\&
			\times\int_{0}^{\gamma_{\text{th},n}}\gamma^{-1}\text{G}^{3,0}_{1,3}\left[\frac{\alpha_{\text{F}_n} \beta_{\text{F}_n}}{A_{0{\text{F}_n}}\delta_{{\kappa_n}}^{\frac{1}{\kappa}}}\gamma^{\frac{1}{\kappa}}\Big{|}^{\xi^2_{\text{F}_n}+1}_{\xi^2_{\text{F}_n},\alpha_{\text{F}_n},\beta_{\text{F}_n}}\right]\nonumber\\&
			\times\Bigg[\left(\frac{\gamma_{\text{th},n}-\gamma}{{\bar{\gamma}}_{\text{T}_n}}\right)^{\frac{\xi_{\text{T}_n}^2}{2}}\Gamma\left(\mathbb{C}^n_{2},\frac{\mathbb{C}^n_{3}}{{{\bar{\gamma}}_{\text{T}_n}}^{\frac{\alpha_n}{2}}}\left(\gamma_{\text{th},n}-\gamma\right)^{\frac{\alpha_n}{2}}\right)\nonumber\\&
			+{\left(\mathbb{C}^n_{3}\right)}^{-\frac{\xi_{\text{T}_n}^2}{\alpha_n}}\Upsilon\left(N_{\text{r}}\mu_n,\frac{\mathbb{C}^n_{3}}{{{\bar{\gamma}}_{\text{T}_n}}^{\frac{\alpha_n}{2}}}\left(\gamma_{\text{th},n}-\gamma\right)^{\frac{\alpha_n}{2}}\right)\Bigg]\text{d}\gamma.
	\end{align}}
	
	Series expansion of the lower incomplete gamma function is given as \cite[(8.354.2)]{gradshteyn2000table}
	{\begin{align}\label{iGamma}
			\Gamma\left(a,bx\right)=\Gamma\left(a\right)-&\sum_{z_1=0}^{\infty}\frac{(-1)^{z_1}}{z_1!\left(a+z_1\right)}\left(bx\right)^{a+z_1}, \nonumber\\&
			~~~~~~~~~~~~\text{for } \left[a \neq 0,-1,-2,...\right].
	\end{align}}
	Substituting (\ref{iGamma}) in (\ref{outMRC2}), we obtain
	{\begin{align}\label{outMRC3}
			&\mathcal{P}^{\text{ST}}_{\gamma_{\text{I}_n}}(\gamma_{\text{th},n})=
			\frac{\xi^2_{\text{F}_n}}{\kappa\Gamma(\alpha_{\text{F}_n})\Gamma(\beta_{\text{F}_n})}\frac{\mathbb{C}^n_{1}}{\xi_{\text{T}_n}^2}\nonumber\\&
			\times\int_{0}^{\gamma_{\text{th},n}}\gamma^{-1}\text{G}^{3,0}_{1,3}\left[\frac{\alpha_{\text{F}_n} \beta_{\text{F}_n}}{A_{0{\text{F}_n}}\delta_{{\kappa_n}}^{\frac{1}{\kappa}}}\gamma^{\frac{1}{\kappa}}\Big{|}^{\xi^2_{\text{F}_n}+1}_{\xi^2_{\text{F}_n},\alpha_{\text{F}_n},\beta_{\text{F}_n}}\right]\nonumber\\&
			\times\Bigg[\left(\frac{\gamma_{\text{th},n}-\gamma}{{\bar{\gamma}}_{\text{T}_n}}\right)^{\frac{\xi_{\text{T}_n}^2}{2}}\bigg(\Gamma\left(\mathbb{C}^n_{2}\right)-\sum_{z_1=0}^{\infty}\frac{(-1)^{z_1}}{z_1!\left(\mathbb{C}^n_{2}+z_1\right)}\nonumber\\&
			\times\left(\frac{\mathbb{C}^n_{3}}{{{\bar{\gamma}}_{\text{T}_n}}^{\frac{\alpha_n}{2}}}\right)^{\mathbb{C}^n_{2}+z_1}\left(\gamma_{\text{th},n}-\gamma\right)^{\frac{\alpha_n}{2}\left(\mathbb{C}^n_{2}+z_1\right)}\bigg)\nonumber\\&
			+{\left(\mathbb{C}^n_{3}\right)}^{-\frac{\xi_{\text{T}_n}^2}{\alpha_n}}\sum_{z_1=0}^{\infty}\frac{(-1)^{z_1}}{z_1!\left(N_{\text{r}}\mu_n+z_1\right)}\left(\frac{\mathbb{C}^n_{3}}{{{\bar{\gamma}}_{\text{T}_n}}^{\frac{\alpha_n}{2}}}\right)^{N_{\text{r}}\mu_n+z_1}\nonumber\\&
			\times\left(\gamma_{\text{th},n}-\gamma\right)^{\frac{\alpha_n}{2}\left(N_{\text{r}}\mu_n+z_1\right)}\Bigg]\text{d}\gamma. 
		\end{align}
		This can be written as
		\begin{align}\label{outMRCmainS1_Appendix}
			\mathbb{I}_1(\gamma_{\text{th},n}) - \mathbb{I}_2(\gamma_{\text{th},n}) +\mathbb{I}_3(\gamma_{\text{th},n}),
	\end{align}}
	%	(\ref{outMRC3}) can be simplified as
	%	\begin{align}\label{outMRCmainS1_Appendix}
		%		\mathcal{P}^{\text{ST}}_{\gamma_{\text{I}_n}}(\gamma_{\text{th},n})&=
		%		%
		%		\mathbb{I}_1(\gamma_{\text{th},n}) - \mathbb{I}_2(\gamma_{\text{th},n}) +\mathbb{I}_3(\gamma_{\text{th},n}),
		%	\end{align}
	where identities $\mathbb{I}_1(\gamma_{\text{th},n})$, $\mathbb{I}_2(\gamma_{\text{th},n})$, and $\mathbb{I}_3(\gamma_{\text{th},n})$ are shown in (\ref{outMRC5}), at the top of the next page.
	
	\begin{figure*}
		\begin{align}	\label{outMRC5}
			\mathbb{I}_1(\gamma_{\text{th},n})&=
			\frac{\xi^2_{\text{F}_n}}{\kappa\Gamma(\alpha_{\text{F}_n})\Gamma(\beta_{\text{F}_n})}\frac{\mathbb{C}^n_{1}\Gamma\left(\mathbb{C}^n_{2}\right)}{\xi_{\text{T}_n}^2{{\bar{\gamma}}_{\text{T}_n}}^{\frac{\xi^2_{\text{T}}}{2}}}\int_{0}^{\gamma_{\text{th},n}}\gamma^{-1}\left(\gamma_{\text{th},n}-\gamma\right)^{\frac{\xi^2_{\text{T}_n}}{2}}\text{G}^{3,0}_{1,3}\left[\frac{\alpha_{\text{F}_n} \beta_{\text{F}_n}}{A_{0\text{F}_n}\delta_{\kappa_n}^{1/\kappa}}\gamma^{\frac{1}{\kappa}}\Big{|}^{\xi^2_{\text{F}_n}+1}_{\xi^2_{\text{F}_n},\alpha_{\text{F}_n},\beta_{\text{F}_n}}\right]\text{d}\gamma,\nonumber\\
			\mathbb{I}_2(\gamma_{\text{th},n})&=\frac{\xi^2_{\text{F}_n}}{\kappa\Gamma(\alpha_{\text{F}_n})\Gamma(\beta_{\text{F}_n})}\frac{\mathbb{C}^n_{1}}{\xi_{\text{T}_n}^2{{\bar{\gamma}}_{\text{T}_n}}^{\frac{\xi^2_{\text{T}}}{2}}}\sum_{z_1=0}^{\infty}\frac{(-1)^{z_1}}{z_1!\left(\mathbb{C}^n_{2}+z_1\right)}\left(\frac{\mathbb{C}^n_{3}}{{{\bar{\gamma}}_{\text{T}_n}}^{\frac{\alpha_n}{2}}}\right)^{\mathbb{C}^n_{2}+z_1}\nonumber\\&
			\times\int_{0}^{\gamma_{\text{th},n}}\gamma^{-1}\left(\gamma_{\text{th},n}-\gamma\right)^{\frac{\xi^2_{\text{T}}}{2}+\frac{\alpha_n}{2}\left(\mathbb{C}^n_{2}+z_1\right)}\text{G}^{3,0}_{1,3}\left[\frac{\alpha_{\text{F}_n} \beta_{\text{F}_n}}{A_{0\text{F}_n}\delta_{\kappa_n}^{1/\kappa}}\gamma^{\frac{1}{\kappa}}\Big{|}^{\xi^2_{\text{F}_n}+1}_{\xi^2_{\text{F}_n},\alpha_{\text{F}_n},\beta_{\text{F}_n}}\right]\text{d}\gamma,\nonumber\\
			\mathbb{I}_3(\gamma_{\text{th},n})&=
			\frac{\xi^2_{\text{F}_n}}{\kappa\Gamma(\alpha_{\text{F}_n})\Gamma(\beta_{\text{F}_n})}\frac{\mathbb{C}^n_{1}{\left(\mathbb{C}^n_{3}\right)}^{-\frac{\xi_{\text{T}_n}^2}{\alpha}}}{\xi_{\text{T}_n}^2}\sum_{z_2=0}^{\infty}\frac{(-1)^{z_2}}{z_2!\left(N_{\text{r}}\mu+z_2\right)}\left(\frac{\mathbb{C}^n_{3}}{{{\bar{\gamma}}_{\text{T}_n}}^{\frac{\alpha_n}{2}}}\right)^{N_{\text{r}}\mu+z_2}\nonumber\\&
			\times\int_{0}^{\gamma_{\text{th},n}}\gamma^{-1}\left(\gamma_{\text{th},n}-\gamma\right)^{\frac{\xi^2_{\text{T}}}{2}+\frac{\alpha_n}{2}\left(N_{\text{r}}\mu+z_2\right)}\text{G}^{3,0}_{1,3}\left[\frac{\alpha_{\text{F}_n} \beta_{\text{F}_n}}{A_{0\text{F}_n}\delta_{\kappa_n}^{1/\kappa}}\gamma^{\frac{1}{\kappa}}\Big{|}^{\xi^2_{\text{F}_n}+1}_{\xi^2_{\text{F}_n},\alpha_{\text{F}_n},\beta_{\text{F}_n}}\right]\text{d}\gamma.
		\end{align}
	\end{figure*}
	Applying \cite[(07.34.21.0085.01)]{wolframe} with mathematical computations, closed-form solutions of the identities $\mathbb{I}_1(\gamma_{\text{th},n})$, $\mathbb{I}_2(\gamma_{\text{th},n})$, and $\mathbb{I}_3(\gamma_{\text{th},n})$ are obtained as given in (\ref{outMRC_Closedform}).
	\vspace{-1em}
	\section{Proof of Diversity Analysis for Combining Method}
	Substituting the high SNR PDF expression of the FSO link (\ref{Asym_FSO_PDF}) and the high SNR CDF expression of the THz link (\ref{Asym_THz_CDF}) in (\ref{Asym_MRC1}), we obtain
	\begin{align}	\label{Asym_ST_appendix}
		&\mathcal{P}^{\text{ST,A}}_{\gamma_{\text{I}_n}}(\gamma_{\text{th},n})\approx
		\frac{\xi^2_{\text{F}_n}}{\kappa\Gamma(\alpha_{\text{F}_n})\Gamma(\beta_{\text{F}_n})}\nonumber\\&
		\times\sum_{p=1}^{3} \frac{\overset{3}{\underset{\underset{q\neq p}{q=1}}\prod} \Gamma(\Psi_{4,q}-\Psi_{4,p})}{\overset{1}{\underset{q=1}\prod}\Gamma(\Psi_{3,q}-\Psi_{4,p})}
		\left(\frac{\alpha_{\text{F}_n} \beta_{\text{F}_n}}{A_{0{\text{F}_n}}}\right)^{\Psi_{4,p}}\left(\frac{1}{\delta_{\kappa_n}}\right)^{\frac{\Psi_{4,p}}{\kappa}} \nonumber\\&
		\times\int_{0}^{\gamma_{\text{th},n}}\gamma^{\frac{\Psi_{4,p}}{\kappa}-1}
		\bigg[\frac{\mathbb{C}^n_{1}\Gamma(\mathbb{C}^n_{2})}{\xi_{\text{T}_n}^2{{\bar{\gamma}}_{\text{T}_n}}^{\frac{\xi_{\text{T}_n}^2}{2}}}\left(\gamma_{\text{th},n}-\gamma\right)^{\frac{\xi_{\text{T}_n}^2}{2}}\nonumber\\&
		-
		\frac{\mathbb{C}^n_{1}{\left(\mathbb{C}^n_{3}\right)}^{\mathbb{C}^n_{2}}}{\mathbb{C}^n_{2}\left(\alpha_n N_{\text{r}}\mu_n\right){\bar{\gamma}}_{\text{T}_n}^{\frac{\alpha_n N_{\text{r}}\mu_n}{2}}}\left(\gamma_{\text{th},n}-\gamma\right)^{\frac{\alpha_n N_{\text{r}}\mu_n}{2}}\bigg]\text{d}\gamma,
	\end{align}
	
	Applying the identity \cite[(3.191.1)]{gradshteyn2000table} in (\ref{Asym_ST_appendix}) with mathematical computations, we obtain
	\begin{align}\label{Asym_ST_appendix1}
		&\mathcal{P}^{\text{ST,A}}_{\gamma_{\text{I}_n}}(\gamma_{\text{th},n})\approx
		\frac{\xi^2_{\text{F}_n}}{\kappa\Gamma(\alpha_{\text{F}_n})\Gamma(\beta_{\text{F}_n})}\nonumber\\&
		\times\sum_{p=1}^{3} \frac{\overset{3}{\underset{\underset{q\neq p}{q=1}}\prod} \Gamma(\Psi_{4,q}-\Psi_{4,p})}{\overset{1}{\underset{q=1}\prod}\Gamma(\Psi_{3,q}-\Psi_{4,p})}\left(\frac{\alpha_{\text{F}_n} \beta_{\text{F}_n}}{A_{0{\text{F}_n}}}\right)^{\Psi_{4,p}}\left(\frac{1}{\delta_{\kappa_n}}\right)^{\frac{\Psi_{4,p}}{\kappa}} \nonumber\\&
		\times\bigg[\frac{\mathbb{C}^n_{1}\Gamma(\mathbb{C}^n_{2})}{\xi_{\text{T}_n}^2{{\bar{\gamma}}_{\text{T}_n}}^{\frac{\xi_{\text{T}_n}^2}{2}}}\left(\gamma_{\text{th},n}\right)^{\frac{\xi_{\text{T}_n}^2}{2}+\frac{\Psi_{4,p}}{\kappa}}\text{B}\left(\frac{\xi_{\text{T}_n}^2}{2}+1,\frac{\Psi_{4,p}}{\kappa}\right)\nonumber\\&
		-\frac{\mathbb{C}^n_{1}{\left(\mathbb{C}^n_{3}\right)}^{\mathbb{C}^n_{2}}}{\mathbb{C}^n_{2}\left(\alpha_n N_{\text{r}}\mu_n\right){\bar{\gamma}}_{\text{T}_n}^{\frac{\alpha_n N_{\text{r}}\mu_n}{2}}}\left(\gamma_{\text{th},n}\right)^{\frac{\alpha_n N_{\text{r}}\mu_n}{2}+\frac{\Psi_{4,p}}{\kappa}}\nonumber\\&
		\times\text{B}\left(\frac{\alpha_n N_{\text{r}}\mu_n}{2}+1,\frac{\Psi_{4,p}}{\kappa}\right)\bigg],
	\end{align}
	where $\text{B}\left(a,b\right)=\int_{0}^{1}z^{a-1}(1-z)^{b-1}\text{d}z$ represents the Beta function.
	
	At high SNRs, the transmit SNR $\bar{\gamma}\rightarrow \infty$. Consequently, the average received SNR of the THz and the FSO links also tend to infinity, i.e, $\bar{\gamma}_{\text{T}_n},\delta_{\kappa_n} \rightarrow \infty$. Hence, considering $\bar{\gamma}_{\text{T}_n}=\Delta_1\bar{\gamma}$ and $\delta_{\kappa_n}=\Delta_2\bar{\gamma}$, where $\Delta_1$ and $\Delta_2$ are constants, (\ref{Asym_ST_appendix1}) can be finally simplified as (\ref{Asym_MRC}).
	
	%%--------------------------------
	\bibliographystyle{IEEEtran}
	\tiny
	\bibliography{Ref_IAB}
	
\end{document}